\documentclass[preprint,onecolumn,superscriptaddress]{revtex4}

\usepackage{amsmath}
\usepackage{latexsym}
\usepackage{amssymb}
\usepackage{pdfpages}
\usepackage{graphicx}
\usepackage{subfigure}
\usepackage[colorlinks=true, citecolor=blue, urlcolor=blue]{hyperref}
\usepackage{float}
\usepackage{epstopdf}
\usepackage{pgffor}
\usepackage{amsfonts}
\usepackage{textcomp}
\usepackage{comment}
\usepackage{braket}
\usepackage{booktabs}
\usepackage{bbm}
\usepackage{multirow}
\usepackage{xcolor}
\definecolor{myurlcolor}{rgb}{0,0,0.8}
\definecolor{mycitecolor}{rgb}{0,0,0.8}
\definecolor{myrefcolor}{rgb}{0,0,0.8}
\usepackage{hyperref}
\usepackage[capitalise]{cleveref}
\hypersetup{colorlinks,
linkcolor=myrefcolor,
citecolor=mycitecolor,
urlcolor=myurlcolor}

\usepackage[draft]{fixme}
\usepackage{amsmath,bbm}%,mathrsfs}
\usepackage{graphicx}
\usepackage{amsfonts}
\usepackage{amssymb}
\usepackage{amsmath, amssymb, amsthm,verbatim,graphicx,bbm}
\usepackage{mathrsfs}
\usepackage{color,xcolor,longtable}

\makeatletter
\usepackage{pifont}
\makeatother

\usepackage{epstopdf}
\usepackage{subfigure}
\usepackage{appendix}

\makeatletter
\usepackage{pifont}
\makeatother

\usepackage{dcolumn}
\usepackage{epstopdf}

\begin{document}

\title{Toward Heisenberg Scaling in Non-Hermitian Metrology at the Quantum Regime}
\author{Xinglei Yu}
\affiliation{School of Physical Science and Technology, Ningbo University, Ningbo, 315211, China}
\author{Xinzhi Zhao}
\affiliation{School of Physical Science and Technology, Ningbo University, Ningbo, 315211, China}
\author{Liangsheng Li}
\affiliation{National Key Laboratory of Scattering and Radiation, Beijing 100854, China}
\author{Xiao-Min Hu}
\affiliation{CAS Key Laboratory of Quantum Information, University of Science and Technology of China, Hefei 230026, China}
\affiliation{CAS Center for Excellence in Quantum Information and Quantum Physics, University of Science and Technology of China, Hefei 230026, China}
\author{Xiangmei Duan}
\affiliation{School of Physical Science and Technology, Ningbo University, Ningbo, 315211, China}
\author{Haidong Yuan}
\email{hdyuan@mae.cuhk.edu.hk}
\affiliation{Department of Mechanical and Automation Engineering, The Chinese University of Hong Kong, Hong Kong}
\author{Chengjie Zhang}
\email{cjzhang@ustc.edu}
\affiliation{School of Physical Science and Technology, Ningbo University, Ningbo, 315211, China}
\affiliation{Hefei National Laboratory, University of Science and Technology of China, Hefei 230088, China}
%\date{\today}
%\maketitle

\begin{abstract}
  Non-Hermitian quantum metrology, an emerging field at the intersection of quantum estimation and non-Hermitian physics, holds promise for revolutionizing precision measurement. Here, we present a comprehensive investigation of non-Hermitian quantum parameter estimation in the quantum regime, with a special focus on achieving Heisenberg scaling. We introduce a concise expression for the quantum Fisher information (QFI) that applies to general non-Hermitian Hamiltonians, enabling the analysis of estimation precision in these systems. Our findings unveil the remarkable potential of non-Hermitian systems to attain the Heisenberg scaling of $1/t$, where $t$ represents time. Moreover, we derive optimal measurement conditions based on the proposed QFI expression, demonstrating the attainment of the quantum Cram\'{e}r-Rao bound. By constructing non-unitary evolutions governed by two non-Hermitian Hamiltonians, one with parity-time symmetry and the other without specific symmetries, we experimentally validate our theoretical analysis. The experimental results affirm the realization of Heisenberg scaling in estimation precision, marking a substantial milestone in non-Hermitian quantum metrology.
\end{abstract}

\date{\today}

\maketitle

\section{\normalsize Introduction}
The assumption of Hamiltonian Hermiticity has long been regarded as a fundamental requirement in quantum mechanics, ensuring real energy eigenvalues and unitary evolution. However, the emergence of research on non-Hermitian Hamiltonians with parity-time ($\mathcal{PT}$) symmetry has challenged this concept and sparked substantial interest. This class of non-Hermitian Hamiltonians exhibits an intriguing property---an entirely real spectrum, especially at exceptional points (EPs), which serve as critical thresholds distinguishing the $\mathcal{PT}$-symmetry broken and unbroken regimes \cite{nh1,nh2}. Over past two decades, there has been plenty of research and developments in various fields based on the interesting features of $\mathcal{PT}$-symmetric non-Hermitian Hamiltonians, such as single-mode lasers \cite{sml1,sml2,sml3}, laser-absorbers \cite{la1,la2,la3}, topological mode transfer \cite{tlet1,tlet2}, metamaterials \cite{memt1,memt3,memt4,memt5} and nonreciprocal device \cite{nrd1,nrd2}, etc. These endeavors have reshaped our understanding and opened up frontiers in quantum science and engineering, leveraging the unique characteristics of non-Hermitian systems with $\mathcal{PT}$ symmetry.

%Among various applications, non-Hermitian quantum metrology attracts growing interest. Enhanced-sensitivity near EP is realized in cavity systems with balanced gain and loss \cite{nqs1,nqs2,nqs3,nqs4,nqs5}. Quantum noise theory of non-Hermitian sensors \cite{nqs6,nqs7} and quantum Fisher information (QFI) in open systems \cite{nqe1,nqe2} were also studied. But the non-Hermitian systems implemented and discussed in these works are still classical (wave) systems, the experiments about quantum non-Hermitian systems have been followed with interest in recent years, for example, single photon network \cite{spn1,spn2,spn3,spn4,spn5,spn6}, cold atoms \cite{ca1,ca2}, trapped ions \cite{ti1,ti2}, superconducting circuit \cite{nhqt1,nhqt2}, and single nitrogen-vacancy center \cite{nv1,nv2}, etc. However, there is still a lack of experiments about the non-Hermitian quantum metrology in quantum regime \cite{spn4}. The previous non-Hermitian metrology works in classical regime consider more about amplitude information \cite{nqs1,nqs2,nqs3,nqs4,nqs5}, experiments of non-Hermitian quantum parameter estimation in quantum regime remains unexplored.
Among various applications, non-Hermitian metrology has emerged as a captivating area of study, attracting considerable interest and attention. Previous studies have focused on achieving enhanced sensitivity near EPs in classical wave systems with balanced gain and loss \cite{nqs1,nqs2,nqs3,nqs4,nqs5}. Additionally, quantum noise theory of non-Hermitian sensors \cite{nqs6,nqs7} and QFI in open systems \cite{nqe1,nqe2} have been investigated. However, most of the implemented and discussed non-Hermitian systems in these works are still in the classical regime, the investigation of non-Hermitian quantum metrology in the quantum regime is still in its early stage.

Recent advancements have showcased promising progress in realizing non-Hermitian quantum systems across various platforms. These developments encompass single photon networks \cite{spn1,spn2,spn3,spn4,spn5,spn6}, cold atoms \cite{ca1,ca2}, trapped ions \cite{ti1,ti2}, superconducting circuits \cite{nhqt1,nhqt2}, and single nitrogen-vacancy centers \cite{nv1,nv2}. Furthermore, important research has demonstrated the evolution of quantum non-Hermitian systems in nuclear magnetic resonance quantum systems \cite{nevo1,ndynamic2}. Non-Hermitian operators have also been investigated as observables for quantum estimation \cite{nhms}. %Interestingly, a new QCRB and non-Hermitian QFI were proposed recently, without the assumption of non-Hermitian system \cite{nhms}. Instead, the non-Hermitian operator is utilized as the measurement, which yields some interesting results, such as a mixed state may exhibit larger QFI than a pure state.
Utilizing the non-Hermitian quantum dynamics for quantum metrology, however, is still a largely unexplored area.

Extensive research has been devoted to achieving the coveted Heisenberg scaling in the realm of Hermitian systems \cite{HL1,HL2,HL3,HL4}. Various strategies have been explored, with the parallel scheme gaining prominence. This approach leverages entangled states as input to achieve Heisenberg precision \cite{entgHL1,entgHL2,entgHL3}. However, the preparation of high-quality, large entangled states poses a substantial challenge. As an alternative, the direct sequential scheme has emerged as another viable avenue to attain Heisenberg scaling without relying on entangled probe states \cite{seqHL1,seqHL2,seqHL3,seqHL4}. Substantial progress has been made in understanding the conditions necessary for achieving Heisenberg scaling in Hermitian systems under both the parallel and sequential schemes \cite{Zhou2018,Rafal2017}. For non-Hermitian systems in the quantum regime, such understanding is still very limited.

In this work, we investigate non-Hermitian quantum metrology both theoretically and experimentally. Firstly, we propose a concise expression for the QFI that is applicable to general non-Hermitian Hamiltonians. This expression allows us to analyze the estimation precision of non-Hermitian systems. Importantly, we find that Heisenberg scaling, characterized by an inverse scaling with time $t^{-1}$, can be achieved in non-Hermitian systems.
Furthermore, we derive the condition for optimal measurements based on the proposed QFI expression. By identifying the optimal measurement strategy, we demonstrate that the estimation precision can reach the fundamental limit characterized by the quantum Cram\'{e}r-Rao bound. To experimentally validate our theoretical findings, we construct a non-unitary evolution governed by a $\mathcal{PT}$-symmetric non-Hermitian Hamiltonian and estimate the associated parameters. By employing the condition for optimal measurements obtained from our theoretical analysis, we achieve estimation precision that matches well with the QCRB. Remarkably, the experimental results reveal that the precision follows Heisenberg scaling $t^{-1}$ for both multiplicative and non-multiplicative Hamiltonians. Our theory is universally applicable and independent of the symmetries of non-Hermitian Hamiltonians. This research not only enriches our understanding of non-Hermitian systems but also opens up exciting avenues for Heisenberg-limited quantum metrology.

\section{\normalsize Resutls}
\subsection{QFI for general non-Hermitian Hamiltonians}
The precision of a quantum system is theoretically limited by the QCRB, as given by $(\Delta\hat{\theta})^2\geq1/(n\mathcal{F}_\theta)$ \cite{qem1,qem2,qem3,qem4,hqem1,hqem2}. Here $\theta$ is the unknown parameter to be estimated, $(\Delta\hat{\theta})^2$ is the variance of an unbiased estimator $\hat{\theta}$, $n$ is the number of the measurements and $\mathcal{F}_\theta$ is QFI that characterizes the optimal estimation precision. For a multiplicative non-Hermitian Hamiltonian $\hat{H}_0=\hat{G}s$, where $\hat{G}$ is the generator and $s$ is the parameter, the evolution of the system is described by the operator $\hat{U}_\theta = e^{-i\hat{H}_0t} = e^{-i\hat{G}\theta}$ \cite{nevo1,nevo2}, where $\theta = st$. If the evolution time $t$ is constant, estimating $\theta$ is equivalent to estimating $s$.

In the case of a pure initial probe state, $|\psi_0\rangle\langle\psi_0|$, the QFI for estimating $\theta$ can be expressed as $\mathcal{F}_\theta = 4(\langle G^\dagger \hat{G}\rangle_\theta - \langle \hat{G}^\dagger\rangle_\theta \langle \hat{G}\rangle_\theta)$ \cite{nhqe1}, where $\langle \hat{G}\rangle_\theta = \bra{\varphi_\theta}\hat{G}\ket{\varphi_\theta}$ represents the expectation value of the normalized output state, and $\ket{\varphi_\theta} = \hat{U}_\theta\ket{\psi_0}/\sqrt{\bra{\psi_0}\hat{U}_\theta^\dagger\hat{U}_\theta\ket{\psi_0}}$ represents the normalized output state \cite{evo1}. We refer the interested readers to the Supplementary Materials for the justification of such normalization. %is performed in this way for the output state based on the reasons presented in the Supplementary Notes.

For general non-multiplicative non-Hermitian Hamiltonians, the QFI cannot be expressed in the form mentioned earlier. In this work, we propose a general expression for the QFI that is applicable to both multiplicative and non-multiplicative non-Hermitian Hamiltonians. Consider a general non-Hermitian Hamiltonian $\hat{H}_\alpha$ which does not necessarily take the multiplicative form, the evolution operator is given by $\hat{U}_\alpha=e^{-i\hat{H}_\alpha t}$. The generator of the parameter $\alpha$ is denoted as $\hat{h}_\alpha=i(\partial_\alpha \hat{U}_\alpha)\hat{U}^{-1}_\alpha$. Based on this, we can write the QFI as
\begin{eqnarray}
\mathcal{F}_\alpha=4(\langle \hat{h}^{\dagger}\hat{h}\rangle_{\alpha}-\langle \hat{h}^{\dagger}\rangle_{\alpha}\langle \hat{h}\rangle_{\alpha}). \label{MQFI}
\end{eqnarray}
This expression provides a more general way to analyze the QFI for non-Hermitian Hamiltonians, regardless of whether they take the multiplicative form or not. One specific application of this expression is the analysis of enhanced or reduced sensitivity near EPs. By utilizing this expression, we can deduce that the sensitivity of quantum sensors is reduced near EPs for two-level multiplicative non-Hermitian Hamiltonians. However, for non-multiplicative non-Hermitian Hamiltonians, the sensitivity may be enhanced and is affected by the modulus of the difference $|\Delta\lambda|$ between two eigenvalues of $\hat{h}$ near the EP. This expression thus opens up avenues for the study of quantum metrology near EPs. Detailed derivations and discussions of Eq. (\ref{MQFI}) can be found in the Supplementary Materials. It should be noted that when the Hamiltonian is Hermitian, this expression reduces to the previously reported form in Refs. \cite{qem3,qem4,hqem2,hqem1} for Hermitian systems.

In addition to the previous analysis, with the expression for the QFI given in Eq.~(\ref{MQFI}), we can further explore whether the Heisenberg scaling can be achieved in non-Hermitian systems using the direct sequential scheme. While the sequential scheme has been well-established as a means to attain Heisenberg scaling of $1/t$ in Hermitian systems, the applicability of this scaling to non-Hermitian systems has remained uncertain due to the previous formulation of the quantum Fisher information (QFI) using the expression $\mathcal{F}_\alpha=4(\langle\partial_\alpha\varphi_\alpha|\partial_\alpha\varphi_\alpha\rangle-|\langle\partial_\alpha\varphi_\alpha|\varphi_\alpha\rangle|^2)$. However, with the expression for the QFI provided in Eq. (\ref{MQFI}), we can now delve into the possibility of achieving the Heisenberg scaling in non-Hermitian systems through the direct sequential scheme. Specifically in the case of multiplicative non-Hermitian Hamiltonian where the evolution operator is given by $\hat{U}=e^{-i\hat{H}_0t}=e^{-i\hat{G}ts}$ (estimating the parameter $s$), the generator of the parameter $s$ is $\hat{G}t$. Utilizing Eq. (\ref{MQFI}), we can calculate the QFI as $\mathcal{F}_s=4t^2(\langle \hat{G}^{\dagger}\hat{G}\rangle_s-\langle \hat{G}^{\dagger}\rangle_s\langle \hat{G}\rangle_s)$. It becomes evident that the precision of the estimation, given by $\sigma(\hat{s})\geq1/\sqrt{\mathcal{F}_s}$, achieves the Heisenberg scaling of $t^{-1}$.

\begin{figure}[b]
\centering
\includegraphics[ width=0.47\textwidth]{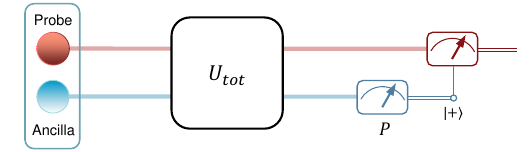}
\caption{Post-selected scheme for the non-unitary evolution. The operator $\hat{U}_{tot}$ is an unitary evolution, we effectively obtain the evolution $\hat{U}'_{PT}=F\hat{U}_{PT}$ of the $\mathcal{PT}$-symmetric Hamiltonian $\hat{H}_{PT}$ for probe qubit after post-selection.  }\label{circuit}
\end{figure}

\begin{figure*}[t]
\centering
\includegraphics[ width=1\textwidth]{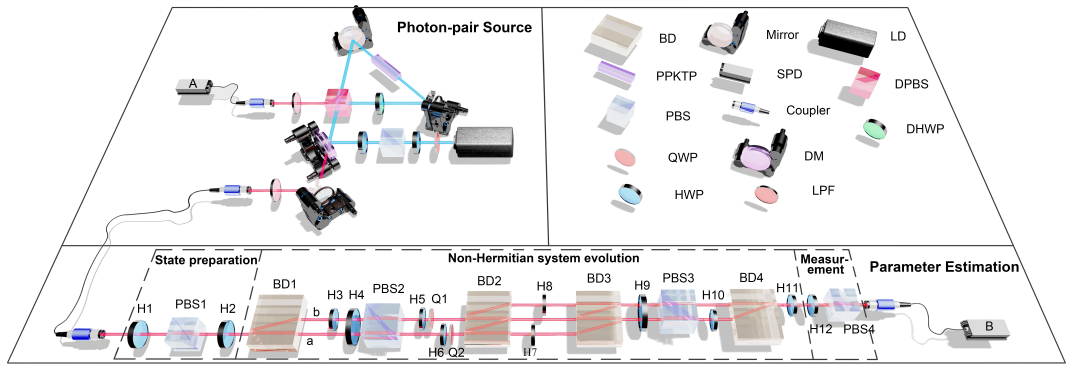}
\caption{Schematic illustration of the experimental setup. Photons pair are generated by a PPKTP crystal, the single target photon is heralded by trigger photon and prepared as probe state. The probe states are purified and rotated respectively by half-wave plate (HWP) and polarization beam splitter (PBS) in the module of state preparation, and then evolve in the module non-Hermitian system evolution, the parameter $\alpha$ is determined by H5 and H6. The output states after evolution are measured by PBS and HWP in the module of measurement.}\label{fig}
\end{figure*}
Achieving the ultimate precision in quantum metrology, as quantified by QCRB, requires the identification of optimal measurement strategies. It is well-known one of the optimal measurement is the projections on the eigenbasis of the symmetric logarithmic derivative (SLD) operator $L_\alpha$ \cite{qem3,qem4,hqem2,hqem1}, which can be obtained as $L_\alpha=2(|\partial_\alpha\varphi_\alpha\rangle\langle\varphi_\alpha|+|\varphi_\alpha\rangle\langle\partial_\alpha\varphi_\alpha|)$ for pure state. However, it is worth noting that the optimal measurement may not be unique, and the computation of the SLD may not be easy using the equation $(\hat{L}_\alpha\tilde{\rho}_\alpha+\tilde{\rho}_\alpha\hat{L}_\alpha)/2=\partial_\alpha\tilde{\rho}_\alpha$, where $\tilde{\rho}_\alpha$ is the normalized output density matrix.

In the pursuit of optimal measurements in non-Hermitian systems, a condition has been proposed in Ref. \cite{nhqe1} for multiplicative Hamiltonians. Here, we further generalize this condition to encompass general non-Hermitian Hamiltonians.  Consider a Hermitian operator $\hat{A}$ as the observable with $\delta\hat{A}=\hat{A}-\langle\hat{A}\rangle_\alpha$ and $(\Delta\hat{A})^2=\langle\delta\hat{A}^\dagger \delta\hat{A}\rangle_\alpha=\langle\hat{A}^\dagger\hat{A}\rangle_\alpha-\langle\hat{A}^\dagger\rangle_\alpha\langle\hat{A}\rangle_\alpha$, the precision of the parameter estimation can be characterized via the error propagation formula $(\Delta\alpha)^2=(\Delta\hat{A})^2/(n|\partial_\alpha\langle\hat{A}\rangle_\alpha|^2)$ \cite{ero1,ero2}.
Using the non-Hermitian uncertainty relationship $(\Delta \hat{A})^2(\Delta \hat{B})^2\geq|\langle\hat{A}^{\dagger}\hat{B}\rangle_\alpha-\langle \hat{A}^{\dagger}\rangle_\alpha\langle\hat{B}\rangle_\alpha|$ \cite{ur1,ur2,ur3,ur4,zx} by taking $\hat{B}$ as $\hat{h}$, we can obtain  $(\Delta \hat{A})^2(\Delta \hat{h})^2\geq |\partial_\alpha \langle \hat{A}\rangle_\alpha|^2/4$ (see Supplementary Materials). This inequality provides a lower bound on the variance of the estimator given by
\begin{eqnarray}
(\Delta\alpha)^2=\frac{(\Delta \hat{A})^2}{n|\partial_\alpha \langle \hat{A}\rangle_\alpha|^2}\geq\frac{(\Delta \hat{A})^2}{4n(\Delta \hat{A})^2(\Delta\hat{h})^2}=\frac{1}{n\mathcal{F}_{\alpha}},
\end{eqnarray}
which is exactly the QCRB. This bound represents the fundamental limit on the precision of parameter estimation. Importantly, the bound is saturated if and only if the observable $\hat{A}$ satisfies
\begin{eqnarray}
\ket{f}=ic\ket{g}, \label{measure}
\end{eqnarray}
where $\ket{f}=\delta\hat{h}\ket{\varphi_\alpha}$, $\ket{g}=\delta\hat{A}\ket{\varphi_\alpha}$, and $c$ is a real number. This condition specifies the relationship between the observables $\hat{A}$ and $\hat{h}$ required to saturate the QCRB. By satisfying the condition in Eq.\;(\ref{measure}), one can attain optimal measurements in non-Hermitian systems, thereby achieving the ultimate precision allowed by the QCRB. This generalized condition opens up possibilities for designing optimal measurement strategies.

\subsection{Model of the experiment system}
For the experiment, we consider a $\mathcal{PT}$-symmetric non-Hermitian Hamiltonian given by
\begin{eqnarray}
\hat{H}_{PT}=s\begin{pmatrix} i\sin\alpha & 1 \\ 1 & -i\sin\alpha  \end{pmatrix},
\end{eqnarray}
where $s$ and $\alpha$ are real parameters. In this case, we assume that the $\mathcal{PT}$-symmetry is not broken, so we have $0<\alpha<\pi/2$. The eigenvalues of $\hat{H}_{PT}$ are $\lambda_\pm=\pm s\cos{\alpha}$, and the corresponding normalized eigenstates are $\ket{\lambda_+}=(e^{i\alpha/2},e^{-i\alpha/2})^T/\sqrt{2}$ and $\ket{\lambda_-}=(e^{-i\alpha/2},-e^{i\alpha/2})^T/\sqrt{2}$.

Furthermore, the non-unitary evolution operator is given by
\begin{eqnarray}
\hat{U}_{PT}&=&e^{-i\hat{H}_{PT}t}\nonumber\\
&=&\frac{1}{\cos{\alpha}}\begin{pmatrix}\cos(ts\cos\alpha-\alpha) & -i\sin(ts\cos\alpha) \\ -i\sin(ts\cos\alpha) & \cos(ts\cos\alpha+\alpha)\end{pmatrix},\quad
\end{eqnarray}
where $t$ represents time. This non-unitary evolution can be accomplished using two-qubit dilated systems, which consist of a probe qubit and an ancilla qubit. The circuit representation of the total evolution $\hat{U}_{tot}$ for the two qubits is illustrated in Fig.\;\ref{circuit}. After post-selecting the ancilla qubit, the probe qubit undergoes a non-unitary evolution. However, it is essential to note that this non-unitary evolution occurs with a certain probability conditioned on the post-selection, which introduces a loss of a portion of states \cite{spn1,spn2,spn3,spn4,qcompt}. Importantly, even when accounting for this loss, the Heisenberg scaling is still achievable.%For more detailed information, please refer to the Supplementary Notes.

\begin{figure}[b]
\centering
\includegraphics[ width=0.49\textwidth]{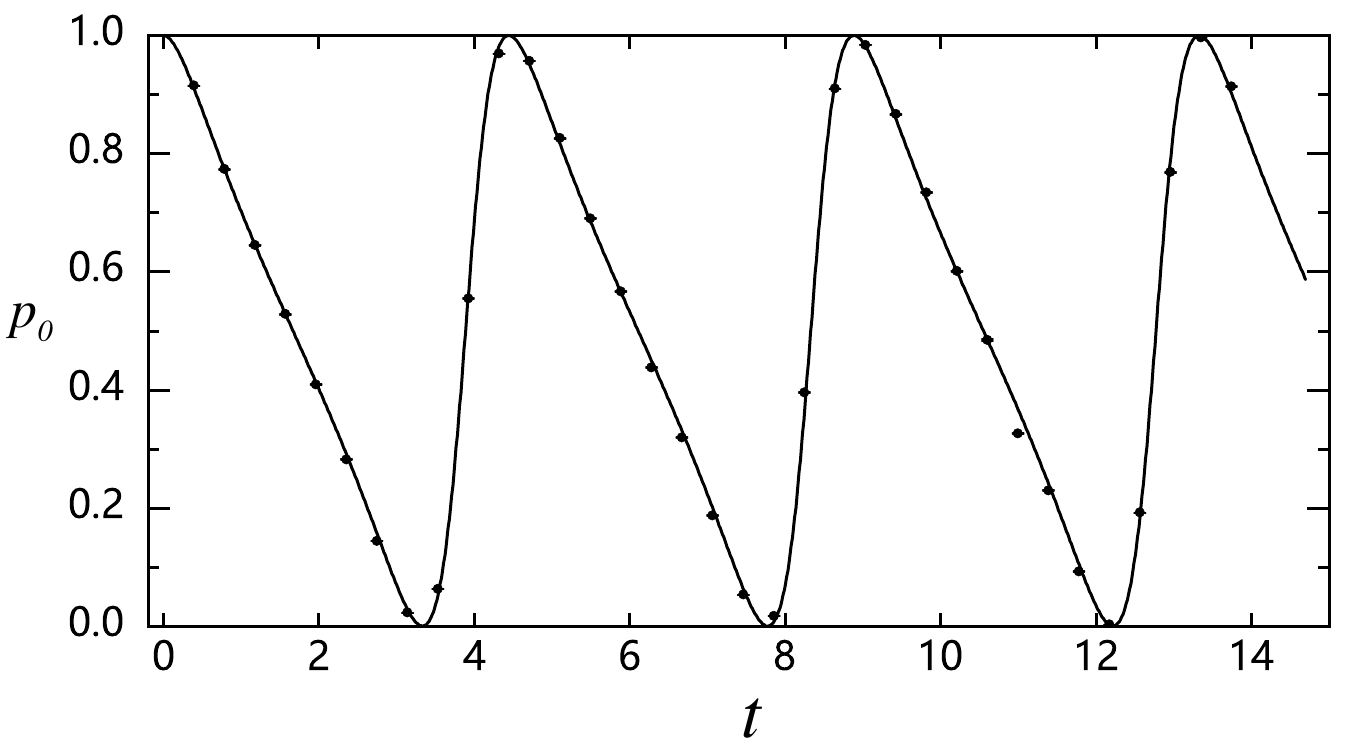}
\caption{The probabilities of measurement outcomes for varying $t$. The black dots represent the experimentally measured data of $p_0$ for varying $t$. And we set $s=1$, $\alpha=\pi/4$, the measurement performed is $\hat{A}=|0\rangle\langle0|$ and the probe state is $\ket{\psi_0}=\ket{0}$. The black solid line is the theoretical value of $p_0=\bra{\varphi}\hat{A}\ket{\varphi}$, the data points match well with the theoretical curve.}\label{nprobability}
\end{figure}

The Heisenberg precision can be achieved for the estimation of $s$ since the Hamiltonian $\hat{H}_{PT}$ is in a multiplicative form with respect to $s$. When the probe state is initially in the state $\ket{\psi_0}=\ket{0}$, the QFI can be obtained as
\begin{equation}\label{multiconditim}
\mathcal{F}_{s}(t)=\frac{4t^2\cos^4\alpha}{[-1+\sin\alpha\sin(\alpha-2st\cos\alpha)]^2}.
\end{equation}
This expression provides the QFI for the parameter $s$ at a general time $t$ which achieves the Heisenberg scaling. For the estimation of $\alpha$ where the non-Hermitian Hamiltonians $\hat{H}_{PT}$ does not take the multiplicative form, we can use Eq.\;(\ref{MQFI}) to obtain the QFI as
\begin{equation}\label{multiconditim}
\mathcal{F}_{\alpha}(t)=\Big[\frac{1-\sec\alpha\cos(\alpha-2st\cos\alpha)+2st\sin\alpha}{\sec\alpha-\sin(\alpha-2st\cos\alpha)\tan\alpha}\Big]^2.
\end{equation}
This also achieves the Heisenberg scaling $t^2$. More detailed analysis and information can be found in the Supplementary Materials.

In addition, to illustrate the generality of Eq.~(\ref{MQFI}), we investigate a broader scenario involving a non-Hermitian Hamiltonian $\hat{H}_\kappa=\kappa|0\rangle\langle1|+|1\rangle\langle0|$ without special symmetries. The experimental precision also achieves Heisenberg scaling, aligning with the theoretical analysis. The details are presented in Supplementary Materials.

\begin{figure*}[t]
\centering
\includegraphics[ width=1\textwidth]{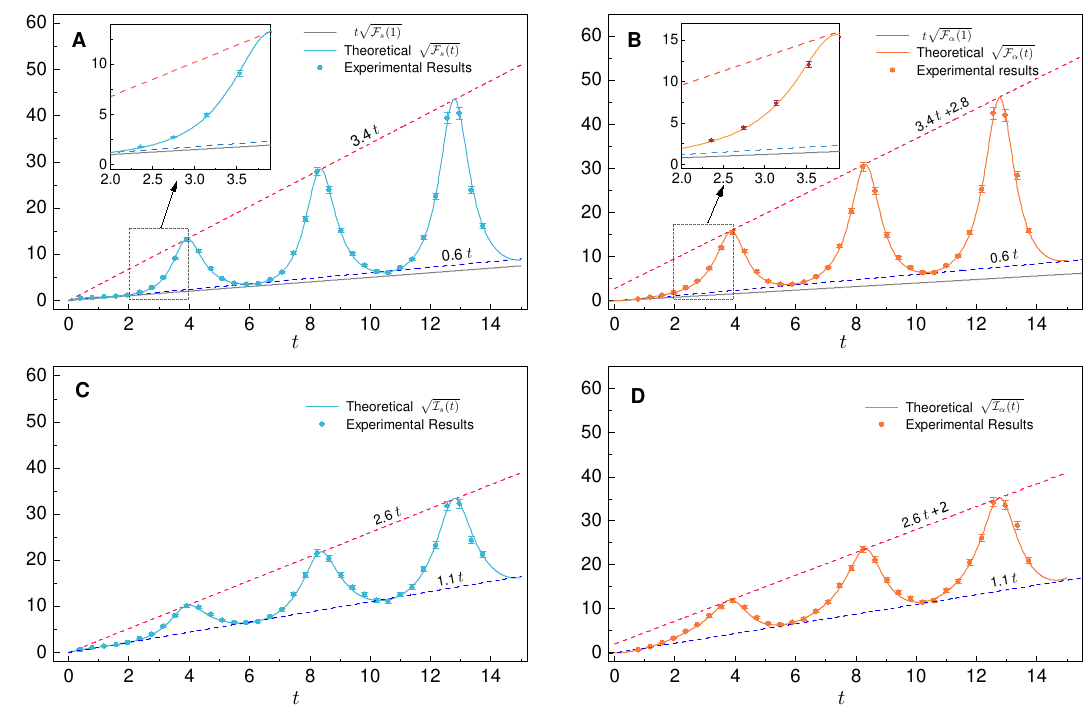}
\caption{QFI for varying time $t$. The probe state is set as $\ket{0}$, the measurement performed is $\hat{A}$, and the condition for optimal measurements is satisfied. The practical value of $s$ and $\alpha$ we set are $1$ and $\pi/4$. (A) The square root of QFI when estimate $s$, the green dots are the experimental data and the green solid line is the theoretical value of $\sqrt{\mathcal{F}_s(t)}$. (B) The square root of QFI when estimate $\alpha$, the orange dots are the experimental data and the orange solid line is the theoretical value of $\sqrt{\mathcal{F}_\alpha(t)}$. (C) The QFI multiplied by normalized coefficient $K_s$. (D) The QFI multiplied by normalized coefficient $K_\alpha$.}\label{HLS}
\end{figure*}

\subsection{Experimental setup and results}

The experimental setup, as depicted in Fig.\;\ref{fig}, consists of four modules:
(1): Photon pair source: A periodically poled potassium titanyl phosphate (PPKTP) crystal is pumped by a 405 nm laser to generate photon pairs through the process of type-\uppercase\expandafter{\romannumeral2} phase-matched spontaneous parametric down-conversion (SPDC) \cite{sps}. One of the photons, called the target photon, serves as the qubit carrier for the non-Hermitian system evolution operation $\hat{U}'_{PT}$. The other photon acts as a trigger signal, referred to as the trigger photon. To ensure data accuracy and reduce environmental interference, we record the coincidence count between photon counter A (trigger photon) and B (target photon), with a coincidence window of 1 ns. (2): State preparation: Photons' polarizations are used to encode states, where a horizontally polarized state $|H\rangle$ corresponds to $|0\rangle$, and a vertically polarized state $|V\rangle$ corresponds to $|1\rangle$. To prepare the target photon, it undergoes a rotation and purification process using half-wave plate 1 (H1) and polarization beam splitter 1 (PBS1), respectively. The target photon is initially prepared in the horizontal state $\ket{0}$ and can be further prepared as an arbitrary linear polarization pure probe state $\ket{\psi_0}=\cos2\phi\ket{0}+\sin2\phi\ket{1}$ with the help of H2. (3): Non-Hermitian system evolution: The non-Hermitian system evolution is achieved by utilizing an ancilla qubit and the projection operation (for post-selection) \cite{spn1}. The detailed construction is described in the Methods section. In practical experiments, the non-unitary evolution is efficiently simulated in an open system by implementing a projection measurement on the ancilla qubit. The actual realized evolution operator is $\hat{U}'_{PT}=F\hat{U}_{PT}$, where $F$ is a function of the estimated parameter. Note that the evolution operator $\hat{U}_{PT}$ multiplied by a function of the estimated parameter does not affect the expression of QFI, and the proof can be found in the Methods section. (4): Measurement: This module consists of HWP and a PBS. Different eigenbases can be used to perform projective measurements. In this case, we choose the measurement $\hat{A}=|0\rangle\langle0|$, which is the optimal measurement when the input probe state is $\ket{0}$. We also experimentally demonstrate the conditions for optimal measurements(see Supplementary Materials for detail).

\begin{figure*}[t]
  \centering
  \includegraphics[ width=1\textwidth]{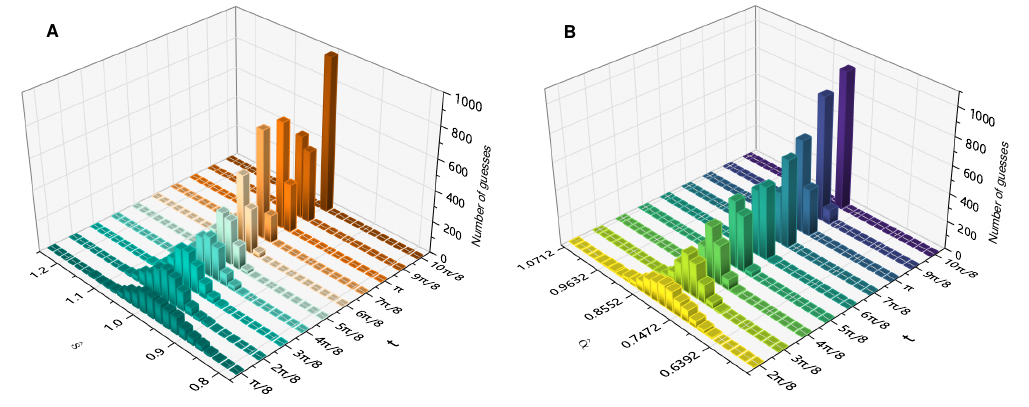}
  \caption{The distribution of estimator $\hat{s}$ and $\hat{\alpha}$ for varying $t$. The distribution becomes more centralized as time $t$ increases (QFI is increased). (A) The distribution of $\hat{s}$. (B) The distribution of $\hat{\alpha}$. }\label{Distb}
\end{figure*}

On the basis of the theoretical results that we discussed earlier, we conducted an experiment to achieve the precision with Heisenberg scaling. In this experiment, we prepared the initial state as $\ket{\psi_0}=\ket{0}$ and estimated the parameters $s$ and $\alpha$ with the optimal measurement $\hat{A}$ for different values of $t$. The true values of the parameters are $s=1$ and $\alpha=\pi/4$. We performed $n=1500\thicksim 2000$ measurements to obtain the probabilities $p_0=\langle\varphi|\hat{A}|\varphi\rangle$ for each evolved probe state, where $\ket{\varphi}$ is the normalized final state. The experimental results, as shown in Fig.\;\ref{nprobability}, are in agreement with the theoretical probabilities. This agreement validates, to a substantial extent, the accuracy of the evolution matrix $\hat{U}_{PT}'$ employed in the experiment.

In order to obtain the statistical information of the estimation, we performed 1000 maximum likelihood estimates and obtained the distributions of the estimators $\hat{s}$ and $\hat{\alpha}$ separately. In Fig.\;\ref{HLS}(A and B), we compare the experimental precision $1/\sigma(\hat{s})$ and $1/\sigma(\hat{\alpha})$ with the theoretical optimal estimation precision $\sqrt{\mathcal{F}_s}$ and $\sqrt{\mathcal{F}_\alpha}$, where $\sigma(\hat{s})$ and $\sigma(\hat{\alpha})$ are the standard deviations of the experimental estimation results. To compare with the theoretical results, we multiplied the coefficient $\sqrt{n}$ by the standard deviation obtained in the experiment. It is important to note that the results shown in Fig.\;\ref{HLS}(A and B) correspond to the precision of a single measurement.
The experimental precision matches well with the theoretical estimation precision for both multiplicative and non-multiplicative Hamiltonians.

In non-Hermitian systems, the estimation precision of successful detection events is characterized by the QFI. However, to determine the ultimate precision for a given resource of probe states, it is necessary to multiply the QFI by the normalization coefficient $K$ of the output state. This can be expressed as $\sqrt{I_s}=\sqrt{K\mathcal{F}_s}$ and $\sqrt{I_\alpha}=\sqrt{K\mathcal{F}_\alpha}$ \cite{nhqe1}. The normalization coefficient $K$ is a periodic function, but it does not impact the overall estimation precision, which still achieves Heisenberg scaling. As shown in Fig.\;\ref{HLS}(C and D), the growth of $\sqrt{I_s}$ and $\sqrt{I_\alpha}$ follows a scaling of $t$, with only a decrease in the oscillation amplitude.

The histograms of parameter estimation results for two estimators are also plotted, as shown in Fig.\;\ref{Distb}(A and B). As indicated in Fig.\;\ref{HLS}, the estimation precision gradually improves within the range of $t$ from $0$ to $10\pi/8$. Consequently, the distribution of the estimator becomes more centralized over time. In Fig.\;\ref{Distb}, it is observed that when $t$ is small, the center of the experimental distribution is noticeably larger than the theoretical value. This discrepancy arises due to the error in the constructed evolution, even though the error itself is relatively small (as shown in Figure \ref{nprobability}). When the QFI is small, even a slight error in the probability of measurement outcomes can result in a notable error in the estimation of the parameter. Additional results can be found in the Supplementary Materials.

\section{\normalsize Discussions and Summary}
One of the main goals in quantum metrology is to achieve the Heisenberg scaling, surpassing the classical limit. Recent studies on systems with Markovian noises have identified the conditions to achieve the Heisenberg scaling\cite{Zhou2018, Rafal2017}. These conditions show that the Heisenberg scaling is not attainable with generic Markovian noises\cite{Rafal2012}. The demonstration of the Heisenberg scaling in non-Hermitian systems presented in this work opens up avenues for identifying systems capable of achieving this scaling. As we have shown, the QFI exhibits an oscillatory behavior as it increases with time in non-Hermitian systems. This phenomenon is parallelled by the periodic oscillation of state distinguishability in non-Hermitian systems\cite{statedisting,spn3}. These oscillations result from the flow of information back from the environment, indicating non-Markovian behavior that exceeds the scope of previous research on achieving the Heisenberg scaling within Markovian dynamics\cite{Zhou2018, Rafal2017}. Although dealing with general non-Markovian systems can be challenging, the presence of non-Markovian behavior in non-Hermitian systems provides possibilities for the identification of systems that can achieve the Heisenberg scaling. By examining the relationship between these oscillatory behaviors and the attainment of the Heisenberg scaling, we anticipate gaining deeper insights into the interplay between quantum metrology and non-Hermitian physics.

In summary, we have introduced a formulation of QFI for general non-Hermitian Hamiltonians, enabling the distinction between systems with enhanced and reduced sensitivity near EPs. This provides a unique perspective for the study of quantum metrology in the vicinity of EPs. We have demonstrated that the Heisenberg scaling can be achieved both theoretically and experimentally in non-Hermitian systems. Additionally, we have derived conditions for optimal measurements, which are applicable to both Hermitian and non-Hermitian systems. Building on this theoretical framework, we have implemented non-unitary evolutions governed by two non-Hermitian Hamiltonians and investigated parameter estimation for such these evolutions. Remarkably, we have achieved the Heisenberg scaling for both multiplicative and non-multiplicative Hamiltonians, with the estimation also reaching the QCRB. The experimental results closely match the theoretical model. It is worth noting that our theory does not make any specific assumptions about the Hamiltonian, and it remains valid for non-Hermitian Hamiltonians without special symmetries. This work represents a notable advancement in both theoretical and experimental research on quantum metrology in non-Hermitian systems.

\section{\normalsize Materials and Methods}
\subsection{Implementation of the non-Hermitian system evolution $\hat{U}'_{PT}$}
The probe state is prepared as $\ket{\psi_0}=\cos2\phi\ket{0}+\sin2\phi\ket{1}$, the photon is separated into two paths by beam displacer 1 (BD1), which introduces the ancilla qubit of path space ($\ket{a}$ represents the path $a$ and $\ket{b}$ represents the path $b$), the horizontal component remains unchanged (path $a$), while the vertical component is deflected into path $b$. The horizontal and vertical components are respectively prepared as $\ket{\varphi_H}=\ket{\psi_H}/\sqrt{\langle\psi_H|\psi_H\rangle}$ and $\ket{\varphi_V}=\ket{\psi_V}/\sqrt{\langle\psi_V|\psi_V\rangle}$ by H5, Q1 ,H6 and Q2, where $\ket{\psi_H}=\hat{U}_{PT}\ket{0}$ and $\ket{\psi_V}=\hat{U}_{PT}\ket{1}$. And finally, $\ket{\varphi_H}$ and $\ket{\varphi_V}$ would be recombined into one path at the output port of the non-Hermitian system evolution, resulting in a loss of photons due to post-selection. As a result, the probe state becomes $F(\cos2\phi\ket{\varphi_H}+\sin2\phi\ket{\varphi_V})$. However, the target output state is $F(\cos2\phi\ket{\psi_H}+\sin2\phi\ket{\psi_V})$, it should be noticed that the gain or loss of two components can be different, i.e., $\langle\psi_H|\psi_H\rangle\neq\langle\psi_V|\psi_V\rangle$, but $\langle\varphi_H|\varphi_H\rangle=\langle\varphi_V|\varphi_V\rangle=1$. To realize it, we add a sub-module consisting of H3, H4 and PBS2, which could control components of $|\varphi_H\rangle$ and $|\varphi_V\rangle$ in two paths. Therefore, before BD2, the probe state is changed to
\begin{eqnarray}
p\cos2\phi\ket{\varphi_H}\ket{a}+q\sin2\phi\ket{\varphi_V}\ket{b},
\end{eqnarray}
where $p=\sin2(\phi_1-\phi_2)$ and $q=\cos2\phi_2$ are controlled by H3 ($\phi_1$) and H4 ($\phi_2$), and $p^2/q^2=\langle\psi_H|\psi_H\rangle/\langle\psi_V|\psi_V\rangle$. The horizontal and vertical components of $\ket{\varphi_H}\ket{a}$ and $\ket{\varphi_V}\ket{b}$ are separated by BD2 and then recombined by H7, H8 and BD3. The post-selection is realized by performing projection operator $\hat{P}=(\ket{a}+\ket{b})(\bra{a}+\bra{b})/2$ on ancilla qubit, and the projection operator is constructed by PBS3 and H9 ($22.5^\circ$). After H10, BD4 and H11, two paths are combined into one path, the output state of probe qubit finally can be written as
\begin{eqnarray}
\hat{U}'_{PT}\ket{\psi_0}&=&\frac{1}{\sqrt{2}}(p\cos2\phi\ket{\varphi_H}+q\sin2\phi\ket{\varphi_V})\nonumber\\
&=&\frac{1}{\sqrt{2}}(\frac{p\cos2\phi\ket{\psi_H}}{\sqrt{\langle\psi_H|\psi_H\rangle}}+\frac{q\sin2\phi\ket{\psi_V}}{\sqrt{\langle\psi_V|\psi_V\rangle}})\nonumber\\
&=&F(\cos2\phi\ket{\psi_H}+\sin2\phi\ket{\psi_V}),
\end{eqnarray}
where $F=p/\sqrt{\langle\psi_H|\psi_H\rangle}=q/\sqrt{\langle\psi_V|\psi_V\rangle}$. The theoretical output state is
\begin{eqnarray}
\hat{U}_{PT}\ket{\psi_0}=\cos2\phi\ket{\psi_H}+\sin2\phi\ket{\psi_V}.
\end{eqnarray}
Therefore, the actual evolution we constructed is $\hat{U}'_{PT}=F\hat{U}_{PT}$.

\subsection{Proof of the invariance of QFI}
We can prove that multiplying the evolution operator with a scalar function, denoted as $F(\alpha)$, does not change the QFI of the normalized state. Let's consider the original expression of QFI with
$\mathcal{F}_\alpha = 4(\langle\partial_{\alpha}\varphi_{\alpha}|\partial_{\alpha}\varphi_{\alpha}\rangle-|\langle\partial_{\alpha}\varphi_{\alpha}|\varphi_{\alpha}\rangle|^{2})$,
where $\alpha$ is the parameter to be estimated. We can decompose the scalar function $F(\alpha)$ into its modulus and phase, $F(\alpha)=R(\alpha)e^{if(\alpha)}$. When the evolution operator is multiplied by $F(\alpha)$, it becomes $\hat{U}'(\alpha)=F(\alpha)\hat{U}(\alpha)=R(\alpha)e^{if(\alpha)}\hat{U}(\alpha)$.
The normalized final state after the multiplication is given by
\begin{eqnarray}
\ket{\varphi'_\alpha}&=&\frac{\hat{U}'(\alpha)\ket{\psi_0}}{\sqrt{\bra{\psi_0}\hat{U}'(\alpha)^\dagger\hat{U}'(\alpha)\ket{\psi_0}}}\nonumber\\
&=&\frac{e^{if(\alpha)}\hat{U}(\alpha)\ket{\psi_0}}{\sqrt{\bra{\psi_0}\hat{U}(\alpha)^\dagger\hat{U}(\alpha)\ket{\psi_0}}}=e^{if(\alpha)}\ket{\varphi_\alpha}.
\end{eqnarray}
It can be observed that if $F(\alpha)$ is a real function, the normalized final state remains unchanged, and consequently, the QFI does not change. However, if $F(\alpha)$ is a complex function, there will be a phase difference $e^{if(\alpha)}$ between $\ket{\varphi_\alpha'}$ and $\ket{\varphi_\alpha}$, and this phase is also a function of $\alpha$.

To simplify the explanation, let's consider this problem from a geometric standpoint. The QFI can also be defined in terms of the Quantum Geometric Tensor (QGT). The QGT, which depends on a set of parameters denoted as $x=(x_1,x_2,\ldots)\in\mathcal{M}$, represents a manifold of the quantum system. The QGT is defined as
$Q_{\mu\nu}(x) = \langle\partial_\mu\varphi(x)|\partial_\nu\varphi(x)\rangle - \langle\partial_\mu\varphi(x)|\varphi(x)\rangle\langle\varphi(x)|\partial_\nu\varphi(x)\rangle$ \cite{qgm1,qgm2,qgm3},
where $\partial_\mu=\partial/\partial x_\mu$, and we have a gauge-invariant metric given by $g_{\mu\nu}=\mathrm{Re}[Q_{\mu\nu}]$. This metric, $g_{\mu\nu}$, remains invariant under gauge transformations of the form $\ket{\varphi'(x)}=e^{if(x)}\ket{\varphi(x)}$. Therefore, the single parameter QFI is exactly the same as the gauge-invariant metric of a one-dimensional manifold $\alpha\in\mathcal{M}$.

According to the gauge invariance, we know that $g_{\mu\nu}$ is invariant under the gauge transformation $\ket{\varphi'(\alpha)}=e^{if(\alpha)}\ket{\varphi(\alpha)}$. Since $\mathcal{F}_\alpha=4\mathrm{Re}[Q_{\alpha\alpha}]=4g_{\alpha\alpha}$, the QFI is also invariant when the evolution operator is multiplied by a function of $\alpha$. \hfill $\square$

\subsection{Analysis of experimental imperfections and details}
In our experimental setup, as depicted in Figure \ref{fig}, the optical path difference between BD1 and BD3 is very small, and H9 is set at an angle of $22.5^\circ$, resulting in a Mach-Zehnder interference. This interference leads to the increase or decrease in the number of photons after post-selection when we input a superposition state of $\ket{0}$ and $\ket{1}$. Consequently, the accuracy of the non-Hermitian evolution is compromised. The fluctuation in the double-coincidence event rate during long-term experiments also affects the accuracy of the evolution. To minimize the interference, it is crucial to maintain a stable experimental environment.

In our experiment, the experimental double-coincidence event rate is approximately 15 kHz after the non-Hermitian system evolution. To obtain the probabilities of measurement outcomes, we measured the final states using both $|0\rangle\langle0|$ and $|1\rangle\langle1|$. We denote the coincidence events of $|0\rangle\langle0|$ and $|1\rangle\langle1|$ as $N_0$ and $N_1$, respectively. The probability of jumping into $\ket{0}$ is calculated as $p_0=N_0/(N_0+N_1)$.

To mitigate the experimental errors caused by the variation in the double-coincidence event rate during long-term experiments, we recorded the coincidence events within a time window of 0.3s. Additionally, we changed the measurement every 500 data points. This approach reduces the fluctuations in the number of measurements between the two different projective operators.

\subsection{Error analysis}
In our experiment, a estimation of the parameter $\alpha$ is based on $n=1500\sim2000$ measurement outcomes. By repeating this $n$ measurements $K=1000$ times, we obtain $1000$ estimation of $\alpha$. Based on this set of estimation results, we could obtain the standard deviation of the estimation $\sigma(\hat{\alpha})$. The error of $\sigma(\hat{\alpha})$, denoted as $\Delta[\sigma(\hat{\alpha})]$ can be approximated by $\Delta[\sigma(\hat{\alpha})]=\sigma(\hat{\alpha})/\sqrt{2(K-1)}$ \cite{standE,seqHL4}. According to QCRB, the experimental QFI depends on the $\sigma(\hat{\alpha})$, so we have $\sqrt{\mathcal{F}_\alpha}=1/[\sigma(\hat{\alpha})\sqrt{n}]$. And the error of $\sqrt{\mathcal{F}_\alpha}$ is approximated by $\Delta(\sqrt{\mathcal{F}_\alpha})=\sqrt{\mathcal{F}_\alpha}/\sqrt{2(K-1)}$, which is employed to draw the error bar in Fig. 4. The error analysis corresponding to parameter $s$ is the same.

%As shown in Fig.\;\ref{fig}, in practical experiment, the optical path difference between the paths of BD1 and BD3 is quite close and H9 is set at $22.5^\circ$, thus forming a Mach-Zehnder interference. Then the number of photons would increase or decrease after post-selection when we input superposition state of $\ket{0}$ and $\ket{1}$, the non-Hermitian evolution is no longer accurate. The variation in double-coincidence event rate during a long-time experiment will also effect the accuracy of evolution. To reduce the influence of the interference, we need to ensure that the experimental environment remains as stable as possible.

%In our experiment, the experimental double-coincidence event rate is $\sim15$ kHz after non-Hermitian system evolution. To get the probabilities of measurement out comes, we measured the final states with both $|0\rangle\langle0|$ and $|1\rangle\langle1|$. We denote the coincidence event of $|0\rangle\langle0|$ and $|1\rangle\langle1|$ respectively as $N_0$ and $N_1$. The probability for jumping into $\ket{0}$ is $p_0=N_0/(N_0+N_1)$. To reduce the experimental errors caused by the variation in double-coincidence event rate during a long-time experiment, the coincidence events within 0.3s is recorded as a data, and we changed the measurement every 500 data. It reduces the fluctuation of number of measurements between the two different projective operators.

\section*{Acknowledgments}
\textbf{Funding:} This work is supported by the National Natural Science Foundation of China (Grant No. 11734015), the Innovation Program for Quantum Science and Technology (Grant No. 2021ZD0301200), and Zhejiang Provincial Natural Science Foundation of China (Grant No. LD24F040001). H.D.Y. acknowledges partial support from the Research Grants Council of Hong Kong (Grants No. 14307420, 14308019,14309022). \textbf{Author contributions:} X.Y. and C.Z. designed and performed the experiment, X.Y., H.Y. and C.Z. contributed to the theoretical analysis, X.Y., X.Z., L.L. and X.-M.H. analyse the theoretical prediction and experimental data.  X.Y., X.D., H.Y. and C.Z. wrote the paper. \textbf{Competing interests:} The authors declare no competing interests. \textbf{Data and materials availability:} The data that support the findings of this study are available within the paper and its Supplementary Information.

\newpage

\renewcommand{\theequation}{S\arabic{equation}}
\renewcommand{\thefigure}{S\arabic{figure}}
\renewcommand{\thetable}{S\arabic{table}}

\section*{\large SUPPLEMENTARY INFORMATION}

\section*{\large \uppercase\expandafter{\romannumeral1}. THEORY}
\subsection*{\normalsize A. Detailed derivation of QFI for general non-Hermitian Hamiltonians}
Consider a non-Hermitian Hamiltonian $\hat{H}_\alpha$ with unknown parameter $\alpha$, the evolution operator can be written as $\hat{U}_\alpha=e^{-i\hat{H}_\alpha t}$ \cite{nevo1}, and the probe state is prepared to be $\rho_0=|\psi_0\rangle\langle\psi_0|$. After evolution, the probe state becomes $\rho_\alpha=\hat{U}_\alpha\rho_0\hat{U}_\alpha^\dagger=|\psi_\alpha\rangle\langle\psi_\alpha|$. When $\alpha$ is changed to $\alpha+d\alpha$, the evolution operator becomes $\hat{U}_{\alpha+d\alpha}\approx\hat{U}_\alpha+\partial_\alpha\hat{U}_\alpha d\alpha$, then we can further deduce the corresponding output state \cite{hqem1},
\begin{eqnarray}
\rho_{\alpha+d\alpha}&\approx&(\hat{U}_\alpha+\partial_\alpha\hat{U}_\alpha d\alpha)\rho_0(\hat{U}_\alpha^\dagger+\partial_\alpha\hat{U}_\alpha^\dagger d\alpha)\nonumber\\
&=&[I+d\alpha(\partial_\alpha\hat{U}_\alpha)\hat{U}_\alpha^{-1}]\hat{U}_\alpha\rho_0\hat{U}_\alpha^\dagger[I+d\alpha(\hat{U}_\alpha^{\dagger})^{-1}(\partial_\alpha\hat{U}_\alpha^\dagger)]\nonumber\\
&=&(I-i\hat{h}_{\alpha} d\alpha )\rho_\alpha(I+i\hat{h}^{\dagger}_{\alpha} d\alpha )\nonumber\\
&\approx&e^{-i\hat{h}_{\alpha}d\alpha}\rho_\alpha e^{i\hat{h}^{\dagger}_{\alpha} d\alpha},
\end{eqnarray}
where
\begin{eqnarray}
\hat{h}_{\alpha}=i(\partial_\alpha\hat{U}_\alpha)\hat{U}_\alpha^{-1} \label{h}
\end{eqnarray}
 is the local generator of the parametric translation of $\hat{U}_\alpha$ with respect to $\alpha$. Sometimes we will omit $\alpha$ when it is clear which parameter is being referred to. 

In non-Hermitian systems, the evolution operator $U_\alpha$ is non-unitary, and as a result, the output state is not normalized after evolution. However, for the measurement process, the probabilities of measurement outcomes still sum up to $1$. Therefore, the output state can be written as \cite{evo1}
\begin{eqnarray}
\tilde{\rho}_\alpha=|\varphi_\alpha\rangle\langle\varphi_\alpha|=\frac{\rho_\alpha}{\mathrm{Tr}\{\rho_\alpha\}}=\frac{\hat{U}_\alpha|\psi_0\rangle\langle\psi_0|\hat{U}_\alpha^\dagger}{\mathrm{Tr}\{\rho_\alpha\}}.
\end{eqnarray}
the corresponding QFI is $\mathcal{F}_\alpha=4(\langle\partial_{\alpha}\varphi_{\alpha}|\partial_{\alpha}\varphi_{\alpha}\rangle-|\langle\partial_{\alpha}\varphi_{\alpha}|\varphi_{\alpha}\rangle|^{2})$. We can rewrite the QFI in terms of $\hat{h}_\alpha$. Let $K_\alpha={\mathrm{Tr}\{\rho_\alpha\}}$, then we have $\ket{\varphi_\alpha}=\hat{U}_\alpha\ket{\psi_0}/\sqrt{K_\alpha}$, the first term $\langle\partial_{\alpha}\varphi_{\alpha}|\partial_{\alpha}\varphi_{\alpha}\rangle$ can be expanded as
\begin{eqnarray}
\langle\partial_{\alpha}\varphi_{\alpha}|\partial_{\alpha}\varphi_{\alpha}\rangle&=&\partial_\alpha(\frac{\bra{\psi_0}\hat{U}_\alpha^\dagger}{\sqrt{K_\alpha}})\partial_\alpha(\frac{\hat{U}_\alpha\ket{\psi_0}}{\sqrt{K_\alpha}})\nonumber\\
&=&\bra{\psi_0}\frac{(\partial_\alpha\hat{U}_\alpha^\dagger)\sqrt{K_\alpha}-\frac{(\partial_{\alpha}K_{\alpha})}{2\sqrt{K_{\alpha}}}\hat{U}_\alpha^\dagger}{K_\alpha}\cdot\frac{\sqrt{K_{\alpha}}(\partial_{\alpha}\hat{U}_{\alpha})-\frac{(\partial_{\alpha}K_{\alpha})}{2\sqrt{K_{\alpha}}}\hat{U}_ {\alpha}}{K_{\alpha}}|\psi_{0}\rangle\nonumber\\
&=&\frac{\bra{\psi_0}(\partial_\alpha\hat{U}_\alpha^\dagger)(\partial_\alpha\hat{U}_\alpha)\ket{\psi_0}}{K_\alpha}-\frac{\partial_\alpha K_\alpha}{2K_\alpha^2}\bra{\psi_0}[\hat{U}_\alpha^\dagger(\partial_\alpha\hat{U}_\alpha)+(\partial_\alpha\hat{U}_\alpha^\dagger)\hat{U}_\alpha]\ket{\psi_0}+\frac{(\partial_\alpha K_\alpha)^2}{4K_\alpha^2}.\nonumber\\
\end{eqnarray}
Similarly, the second term $|\langle\partial_{\alpha}\varphi_{\alpha}|\varphi_{\alpha}\rangle|^{2}$ can be expanded as
\begin{eqnarray}
&&|\langle\partial_{\alpha}\varphi_{\alpha}|\varphi_{\alpha}\rangle|^{2}\nonumber\\
&=&\bigg|\bra{\psi_0}\frac{(\partial_\alpha\hat{U}_\alpha^\dagger)\sqrt{K_\alpha}-\frac{\partial_\alpha K_\alpha}{2\sqrt{K_\alpha}}\hat{U}_\alpha^\dagger}{K_\alpha}\frac{\hat{U}_\alpha}{\sqrt{K_\alpha}}\ket{\psi_0}\bigg|^2=\bigg|\bra{\psi_0}\frac{(\partial_\alpha\hat{U}_\alpha^\dagger)\hat{U}_\alpha}{K_\alpha}\ket{\psi_0}-\frac{\partial_\alpha K_\alpha}{2K_\alpha}\bigg|^2\nonumber\\
&=&\frac{\bra{\psi_0}(\partial_\alpha\hat{U}_\alpha^\dagger)\hat{U}_\alpha\ket{\psi_0}\bra{\psi_0}\hat{U}_\alpha^\dagger(\partial_\alpha\hat{U}_\alpha)\ket{\psi_0}}{K_\alpha^2}-\frac{\partial_\alpha K_\alpha}{2K_\alpha^2}\bra{\psi_0}[\hat{U}_\alpha^\dagger(\partial_\alpha\hat{U}_\alpha)+(\partial_\alpha\hat{U}_\alpha^\dagger)\hat{U}_\alpha]\ket{\psi_0}\nonumber\\
&&+\frac{(\partial_\alpha K_\alpha)^2}{4K_\alpha^2}.
\end{eqnarray}
Collecting this two results, we obtain
\begin{eqnarray}
\mathcal{F}_\alpha&=&4(\langle\partial_{\alpha}\varphi_{\alpha}|\partial_{\alpha}\varphi_{\alpha}\rangle-|\langle\partial_{\alpha}\varphi_{\alpha}|\varphi_{\alpha}\rangle|^{2})\nonumber\\
&=&\frac{\bra{\psi_0}(\partial_\alpha\hat{U}_\alpha^\dagger)(\partial_\alpha\hat{U}_\alpha)\ket{\psi_0}}{K_\alpha}-\frac{\bra{\psi_0}(\partial_\alpha\hat{U}_\alpha^\dagger)\hat{U}_\alpha\ket{\psi_0}\bra{\psi_0}\hat{U}_\alpha^\dagger(\partial_\alpha\hat{U}_\alpha)\ket{\psi_0}}{K_\alpha^2}\nonumber\\
&=&\frac{\bra{\psi_\alpha}(\hat{U}_\alpha^\dagger)^{-1}(\partial_\alpha\hat{U}_\alpha^\dagger)(\partial_\alpha\hat{U}_\alpha)\hat{U}_\alpha^{-1}\ket{\psi_\alpha}}{K_\alpha}-\frac{\bra{\psi_\alpha}(\hat{U}_\alpha^\dagger)^{-1}(\partial_\alpha\hat{U}_\alpha^\dagger)\ket{\psi_\alpha}\bra{\psi_\alpha}(\partial_\alpha\hat{U}_\alpha)\hat{U}_\alpha^{-1}\ket{\psi_\alpha}}{K_\alpha^2}\nonumber\\
&=&\bra{\varphi_\alpha}(\hat{U}_\alpha^\dagger)^{-1}(\partial_\alpha\hat{U}_\alpha^\dagger)(\partial_\alpha\hat{U}_\alpha)\hat{U}_\alpha^{-1}\ket{\varphi_\alpha}-\bra{\varphi_\alpha}(\hat{U}_\alpha^\dagger)^{-1}(\partial_\alpha\hat{U}_\alpha^\dagger)\ket{\varphi_\alpha}\bra{\varphi_\alpha}(\partial_\alpha\hat{U}_\alpha)\hat{U}_\alpha^{-1}\ket{\varphi_\alpha}.\nonumber\\
\end{eqnarray}
This can be expressed in terms of $\hat{h}_{\alpha}=i(\partial_\alpha\hat{U}_\alpha)\hat{U}_\alpha^{-1}$ as
\begin{eqnarray}
\mathcal{F}_\alpha=4(\langle \hat{h}^{\dagger}_{\alpha} \hat{h}_{\alpha}\rangle_\alpha-\langle \hat{h}^{\dagger}_{\alpha}\rangle_\alpha\langle \hat{h}_{\alpha}\rangle_\alpha), \label{QFI}
\end{eqnarray}
where $\langle\bullet\rangle_\alpha=\bra{\varphi_\alpha}\bullet\ket{\varphi_\alpha}$. 

When the Hamiltonian is multiplicative, $\hat{H}=G\theta$, and the evolution operator is given by  $\hat{U}=e^{-iG\theta t}=e^{-iG\alpha}$, where $\theta$ is the parameter which we aim to estimate and $\alpha=\theta t$. If the evolution time $t$ is a constant then estimating $\alpha$ is equivalent to estimate $\theta$. In this case, Equation (\ref{QFI}) takes the form: $\mathcal{F}_\alpha=4(\langle G^\dagger G\rangle_\alpha-\langle G^\dagger\rangle_\alpha\langle G\rangle_\alpha)$ \cite{nhqe1}.

The key to calculating QFI is the local generator $\hat{h}_{\alpha}$, the parametric translation of $\hat{U}_\alpha$. However, in general, it is difficult to directly derive $\hat{h}_{\alpha}$, because $\partial_\alpha\hat{H}_\alpha$ does not commute with $\hat{U}_\alpha$ when calculating $\partial_\alpha\hat{U}_\alpha$. Fortunately, this problem has been addressed in previous works\cite{mth1}, and the generator $\hat{h}_{\alpha}$ can be expressed as \cite{hqem1}
\begin{eqnarray}
\hat{h}_{\alpha}=i(\partial_\alpha\hat{U}_\alpha)\hat{U}_\alpha^{-1}=\int_0^te^{-i\mu H_\alpha}\partial_\alpha H_\alpha e^{i\mu H_\alpha}d\mu.
\end{eqnarray}

Notably, QFI represents the ultimate precision for a single measurement. However, in non-Hermitian systems, there is gains and losses during the non-unitary evolutions. Relying solely on QFI is insufficient to characterize the ultimate estimation precision when considering fixed resources of probe states. Considering the effect of gains and losses on precision, we multiply the QFI by the normalization coefficient $K_\alpha$ of the output state
\begin{eqnarray}
I_\alpha=K_\alpha\mathcal{F}_\alpha.
\end{eqnarray}
This quantity actually characterized the estimation precision for evolution $\hat{U}_\alpha$ in the case of fixed resources of probe states.

Here, we further discuss the precision scaling of a $\mathcal{PT}$-symmetric system $\hat{H}_{PT}$ which is embedded into a larger Hermitian system \cite{statedisting,nevo2}. It is well-known that some $\mathcal{PT}$-symmetric Hamiltonians are pseudo-Hermitian, we can find an invertible Hermitian operator $\hat{\eta}$ that satisfies $\hat{\eta}\hat{H}_{PT}=\hat{H}^\dagger_{PT}\hat{\eta}$. Here, we define $c:=\sum_{i=1}^{N}1/\lambda_i$ and $\hat{\zeta}=c\hat{\eta}-\hat{I}$, where $\lambda_i$ represents the $i$th eigenvalue of $\hat{\eta}$ and $\hat{I}$ is the identity operator. We consider the case of $\mathcal{PT}$-symmetry unbroken, so $\hat{\eta}$ and $\hat{\zeta}$ are positive. The two-qubit dilation Hermitian system can be written as \cite{statedisting,nevo2}
\begin{eqnarray}
\hat{H}_{tot}=\hat{I}\otimes\hat{H}_s+\hat{\sigma}_y\otimes\hat{V},
\end{eqnarray}
where $\hat{H}_s$ and $\hat{V}$ are Hermitian and satisfy
\begin{eqnarray}
\hat{H}_{s}-i\hat{V}\hat{\zeta}^{1/2}=\hat{H}_{PT},\quad \hat{H}_{s}+i\hat{V}\hat{\zeta}^{-1/2}=\hat{\zeta}^{1/2}\hat{H}_{PT}\hat{\zeta}^{-1/2}.
\end{eqnarray}
Solving the above equation, we obtain
\begin{eqnarray}
\hat{H}_s=(\hat{H}_{PT}\hat{\zeta}^{-1/2}+\hat{\zeta}^{1/2}\hat{H}_{PT})(\hat{\zeta}^{1/2}+\hat{\zeta}^{-1/2})^{-1},\\ \hat{V}=i(\hat{H}_{PT}-\hat{\zeta}^{1/2}\hat{H}_{PT}\hat{\zeta}^{-1/2})(\hat{\zeta}^{1/2}+\hat{\zeta}^{-1/2})^{-1}.
\end{eqnarray}
An entangled state $|\Psi_{tot}(t)\rangle$ in two-qubit dilation Hermitian system can be written as
\begin{eqnarray}
|\Psi_{tot}(t)\rangle=|0\rangle\otimes|\psi_{PT}(t)\rangle+|1\rangle\otimes\big(\zeta^{1/2}|\psi_{PT}(t)\rangle\big),
\end{eqnarray}
where $|\psi_{PT}(t)\rangle=e^{-i\hat{H}_{PT}t}|\psi_0\rangle$ is the state evolves in $\mathcal{PT}$-symmetric system and $|\psi_0\rangle$ is the initial state of $|\psi_{PT}(t)\rangle$. It can be proved that $|\Psi_{tot}(t)\rangle$ satisfies the Sch\"{o}rdinger equation of $\hat{H}_{tot}$, we have
\begin{eqnarray}
i\frac{d}{dt}|\Psi_{tot}(t)\rangle&=&i\big[|0\rangle\otimes\frac{d}{dt}|\psi_{PT}(t)\rangle+|1\rangle\otimes(\zeta^{1/2}\frac{d}{dt}|\psi_{PT}(t)\rangle)\big]\nonumber\\
&=&|0\rangle\otimes\hat{H}_{PT}|\psi_{PT}(t)\rangle+|1\rangle\otimes(\zeta^{1/2}\hat{H}_{PT}|\psi_{PT}(t)\rangle)\nonumber\\
&=&|0\rangle\otimes\big(\hat{H}_{s}-i\hat{V}\hat{\zeta}^{1/2}\big)|\psi_{PT}(t)\rangle+|1\rangle\otimes(\hat{H}_{s}+i\hat{V}\hat{\zeta}^{-1/2})(\hat{\zeta}^{1/2}|\psi_{PT}(t)\rangle)\nonumber\\
&=&\Big[\hat{I}\otimes\hat{H}_s+\hat{\sigma}_y\otimes\hat{V}\Big]\Big[|0\rangle\otimes|\psi_{PT}(t)\rangle+|1\rangle\otimes(\zeta^{1/2}|\psi_{PT}(t)\rangle)\Big]\nonumber\\
&=&\hat{H}_{tot}|\Psi_{tot}(t)\rangle.
\end{eqnarray}
Noticing that the $\hat{\eta}$ inner product $\langle\psi_{PT}(t)|\hat{\eta}|\psi_{PT}(t)\rangle=\langle\psi_{0}|\hat{\eta}|\psi_{0}\rangle$ is invariant under the $\mathcal{PT}$ dynamics, so the norm of $|\Psi_{tot}(t)\rangle$ is also invariant $\langle\Psi_{tot}(t)|\Psi_{tot}(t)\rangle=c\langle\psi_{0}|\hat{\eta}|\psi_{0}\rangle$. And the normalized state of $|\Psi_{tot}(t)\rangle$ can be written as $|\psi_{tot}(t)\rangle=|\Psi_{tot}(t)\rangle/\sqrt{c\langle\psi_{0}|\hat{\eta}|\psi_{0}\rangle}$. Therefore, we can obtain $|\psi_{PT}(t)\rangle/\sqrt{c\langle\psi_{0}|\hat{\eta}|\psi_{0}\rangle}$ by performing a projective operator $\hat{P}_0\otimes\hat{I}=|0\rangle\langle0|\otimes\hat{I}$ on an entangled state $|\psi_{tot}(t)\rangle$ which evolves in a larger Hermitian system. In other words, we realize the evolution $\hat{U}^\prime_{PT}=\hat{U}_{PT}/\sqrt{c\langle\psi_{0}|\hat{\eta}|\psi_{0}\rangle}$ of $\mathcal{PT}$-symmetric system $\hat{H}_{PT}$ in a larger Hermitian system via post-selection.

As we discussed in the main text, multiplying the evolution operator with a scalar function does not change $\mathcal{F}_\alpha$, which is the QFI calculated from $\hat{U}_{PT}$. However, for fixed resource of probe states, the scalar function characterizes the probability of realizing $\hat{U}_{PT}$, which will affect the ultimate estimation precision. Therefore, the ultimate estimation precision is actually limited by $K_\alpha\mathcal{F}_\alpha/\big(c\langle\psi_{0}|\hat{\eta}|\psi_{0}\rangle\big)$. The coefficient $c\langle\psi_{0}|\hat{\eta}|\psi_{0}\rangle$ is independent of time $t$. Obviously, the estimation precision scaling for the evolution $\hat{U}_{PT}$ is the same as that for the evolution $\hat{U}^\prime_{PT}$.

\subsection*{\normalsize B. Analysis of QFI for two level non-Hermitian Hamiltonians near EP}
Consider a two-level non-Hermitian operator $\hat{O}$ with two nondegenerate eigenvalues $\lambda_{1,2}$ and two linearly independent eigenvectors $\ket{\lambda_{1,2}}$. Suppose we have an arbitrary normalized state which is able to be expressed by the superposition of eigenstates
\begin{eqnarray}
\ket{\psi}=a\ket{\lambda_1}+b\ket{\lambda_2},
\end{eqnarray}
where $a$ and $b$ are complex numbers, then we define the variance of non-Hermitian operator $\hat{O}$ as $(\Delta\hat{O})^2=\bra{\psi}\hat{O}^\dagger\hat{O}\ket{\psi}-\bra{\psi}\hat{O}^\dagger\ket{\psi}\bra{\psi}\hat{O}\ket{\psi}$. We can further rewrite it as follow
\begin{eqnarray}
(\Delta\hat{O})^2=\bra{\psi}\hat{O}^\dagger\hat{O}\ket{\psi}-\bra{\psi}\hat{O}^\dagger\ket{\psi}\bra{\psi}\hat{O}\ket{\psi}=\bra{\psi}\hat{O}^\dagger(I-|\psi\rangle\langle\psi|)\hat{O}\ket{\psi}=\bra{\psi}\hat{O}^\dagger\hat{P}\hat{O}\ket{\psi}.\nonumber\\
\end{eqnarray}
We can see that for two-level systems, the projective operator $\hat{P}=I-|\psi\rangle\langle\psi|=|\phi\rangle\langle\phi|$ is exactly the pure state that orthogonal to $\ket{\psi}$. Assume $\ket{\phi}=c\ket{\lambda_1}+d\ket{\lambda_2}$, then we have
\begin{eqnarray}
\langle\phi|\psi\rangle=ac^\ast+bd^\ast+ad^\ast\langle\lambda_2|\lambda_1\rangle+bc^\ast\langle\lambda_1|\lambda_2\rangle=0, \label{orthg}
\end{eqnarray}
and $\hat{O}\ket{\psi}$ can be written as $\hat{O}\ket{\psi}=a\lambda_1\ket{\lambda_1}+b\lambda_2\ket{\lambda_2}$. Therefore, the variance can be further written as
\begin{eqnarray}
(\Delta\hat{O})^2=\bra{\psi}\hat{O}^\dagger\hat{P}\hat{O}\ket{\psi}=|\bra{\phi}\hat{O}\ket{\psi}|^2=\big|ac^\ast\lambda_1+bd^\ast\lambda_2+ad^\ast\lambda_1\langle\lambda_2|\lambda_1\rangle+bc^\ast\lambda_2\langle\lambda_1|\lambda_2\rangle\big|^2.\nonumber\\
\end{eqnarray}
According to Eq. (\ref{orthg}), we have
\begin{eqnarray}\label{QFIdifference}
(\Delta\hat{O})^2&=&\big|ac^\ast\lambda_1+bd^\ast\lambda_2+ad^\ast\lambda_1\langle\lambda_2|\lambda_1\rangle-\lambda_2\big(ac^\ast+bd^\ast+ad^\ast\langle\lambda_2|\lambda_1\rangle\big)\big|^2\nonumber\\
&=&\big|ac^\ast(\lambda_1-\lambda_2)+ad^\ast\langle\lambda_2|\lambda_1\rangle(\lambda_1-\lambda_2)\big|^2\nonumber\\
&=&\big|a\big|^2\cdot\big|(\lambda_1-\lambda_2)\big|^2\cdot\big|c^\ast+d^\ast\langle\lambda_2|\lambda_1\rangle\big|^2.
\end{eqnarray}
Obviously, for a given state $\ket{\psi}=a\ket{\lambda_1}+b\ket{\lambda_2}$, the variance $(\Delta\hat{O})^2$ is determined by the modulus of difference between two eigenvalues $\lambda_1$ and $\lambda_2$. As we discussed in section A, for a multiplicative pseudo-Hermitian Hamiltonian $\hat{H}=Gs$, the QFI is the variance of the $G$ which has the same properties with $\hat{H}$. It is well-known that for pseudo-Hermitian Hamiltonians, the eigenvalues gradually degenerate as EP is approached. Therefore, for a given state $\ket{\psi}=a\ket{\lambda_1}+b\ket{\lambda_2}$, the QFI of multiplicative pseudo-Hermitian Hamiltonian $\hat{H}$ would tends to zero near EP.

However, as for non-multiplicative Hamiltonians, QFI is the variance of the generator $h$, the properties of $h$ near EP could be much different from that of Hamiltonian. For example, consider a non-Hermitian Hamiltonian $\hat{H}_1=I\sin\alpha \sigma_z+\cos\alpha\sigma_x$, $\alpha$ is the unknown parameter, the evolution operator is $U_1=e^{-iH_1t}$. Then we can obtain the eigenvalues of local generator $h_1$
\begin{eqnarray}
\lambda_1&=&-\sec2\alpha\sqrt{\frac{\cos(2t\sqrt{\cos2\alpha})+t^2\sin2\alpha\sin4\alpha-1}{2}},\\
\lambda_2&=&\sec2\alpha\sqrt{\frac{\cos(2t\sqrt{\cos2\alpha})+t^2\sin2\alpha\sin4\alpha-1}{2}}.
\end{eqnarray}
%\renewcommand{\arraystretch}{3}
%\begin{eqnarray}
%&&h_1=\nonumber\\
%&&\begin{pmatrix}\frac{i}{2}\cos\alpha\sec^22\alpha[\sqrt{\cos2\alpha}\sin(2\sqrt{\cos2\alpha})-4\cos2\alpha\sin^2\alpha] & -\sec2\alpha[2\cos^2\alpha\sin\alpha+\sin^2(\sqrt{\cos2\alpha})]+\frac{\sin\alpha\sin(2\sqrt{\cos2\alpha})}{2\cos^{3/2}2\alpha} \\ -i & \frac{-i}{2}\cos\alpha\sec^22\alpha[\sqrt{\cos2\alpha}\sin(2\sqrt{\cos2\alpha})-4\cos2\alpha\sin^2\alpha]\end{pmatrix}.\nonumber\\
%\end{eqnarray}
In the case of $\mathcal{PT}$-symmetry unbroken, if $t\gg1$, the modulus of difference $|\Delta\lambda|=|\lambda_2-\lambda_1|$ increases as EP ($\alpha=\pi/4$) is approached. Thus, the QFI of non-multiplicative Hamiltonians could increase near EP. Noticed that the above discussion applies to the case that EP is not exactly reached. Because two eigenstates would coalesce at EP, the state $\ket{\psi}=a\ket{\lambda_1}+b\ket{\lambda_2}$ cannot be able to represent an arbitrary state anymore, it is just the eigenstate.

\begin{figure*}[t]
\centering
\includegraphics[width=0.8\textwidth]{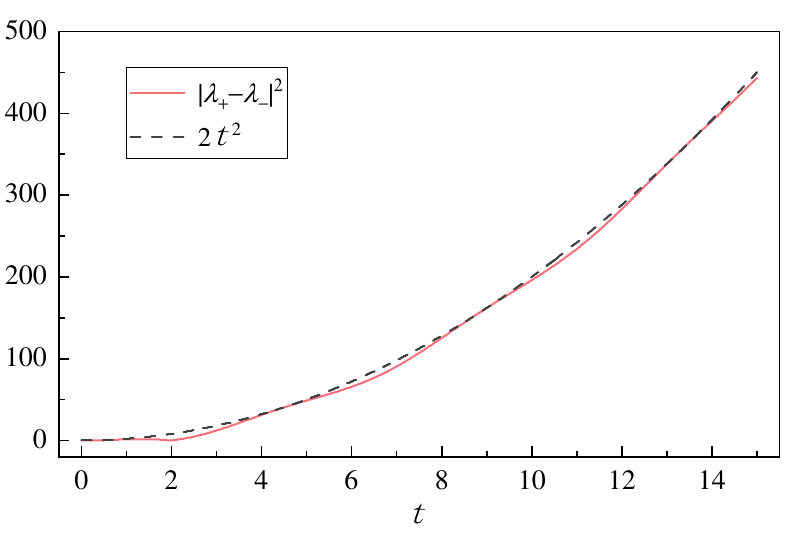}
\caption{The evolution of the modulus of difference $|\lambda_+-\lambda_-|^2$ as function of time $t$.}\label{eigenvalue}
\end{figure*}

Furthermore, we are also able to analyse that whether we can achieve Heisenberg precision for multiplicative non-Hermitian Hamiltonians. For example, the generator of the parameter $\alpha$ for general time $t$ of the non-Hermitian Hamiltonian $\hat{H}_{PT}(\alpha)$ in the main text is
\small{
\begin{eqnarray}
\hat{h}_\alpha(t)
=\frac{\sec\alpha}{2}\begin{pmatrix} i[\sec\alpha+\sin(2st\cos\alpha)-2st\sin^2\alpha] &\quad \sec\alpha\cos(\alpha-2st\cos\alpha)-2st\sin\alpha-1 \\ \sec\alpha\cos(\alpha+2st\cos\alpha)+2st\sin\alpha-1 & \quad -i[\sec\alpha+\sin(2st\cos\alpha)-2st\sin^2\alpha]\end{pmatrix},\nonumber\\
\end{eqnarray}}
\normalsize{
and we can obtain the eigenvalues of $\hat{h}_\alpha(t)$ as follows
\begin{eqnarray}
\lambda_\pm=\pm\frac{\sec\alpha\sqrt{4\cos(2st\cos\alpha)-4+s^2t^2(1-\cos4\alpha)}}{2\sqrt{2}}.
\end{eqnarray}}
As shown in Fig. \ref{eigenvalue}, the growth of the modulus of difference $|\lambda_+-\lambda_-|$ reaches the scale of $t$. According to Eq. (\ref{QFIdifference}), we can derive that the growth of of QFI reaches $t^2$, i.e., Heisenburg limit, and we obtain the exact QFI for general time $t$ when probe state is $\ket{\psi_0}=\ket{0}$ as follows
\begin{eqnarray}
\mathcal{F}_\alpha(t)=\Big[\frac{1-\sec\alpha\cos(\alpha-2st\cos\alpha)+2st\sin\alpha}{\sec\alpha-\sin(\alpha-2st\cos\alpha)\tan\alpha}\Big]^2.
\end{eqnarray}
Obviously, there is $t^2$ in $\mathcal{F}_\alpha(t)$.

\subsection*{\normalsize C. Proof of the condition for optimal measurements}

Based on the expression of QFI we proposed, we give the detailed proof of the condition for optimal measurement in this section. Consider an arbitrary Hermitian operator $\hat{A}$ as measurement operator. According to the error-propagation formula \cite{ero1,ero2}, we have
\begin{eqnarray}
(\Delta\alpha)^2=\frac{(\Delta \hat{A})^2}{n|\partial_\alpha \langle \hat{A}\rangle_\alpha|^2}=\frac{(\Delta\hat{A})^2}{n|\bra{\partial_{\alpha}\varphi_{\alpha}}\hat{A}\ket{\varphi_{\alpha}}+\bra{\varphi_\alpha}\hat{A}\ket{\partial_\alpha\varphi_{\alpha}}|^2},
\end{eqnarray}
where $\ket{\varphi_\alpha}$ is the normalized output state after evolution. For simplicity, let $Q=\bra{\partial_{\alpha}\varphi_{\alpha}}\hat{A}\ket{\varphi_{\alpha}}+\bra{\varphi_\alpha}\hat{A}\ket{\partial_\alpha\varphi_{\alpha}}$ and expand it as follow
\begin{eqnarray}
Q&=&\bra{\partial_{\alpha}\varphi_{\alpha}}\hat{A}\ket{\varphi_{\alpha}}+\bra{\varphi_\alpha}\hat{A}\ket{\partial_\alpha\varphi_{\alpha}}=\bra{\psi_0}\partial_\alpha\Big(\frac{\hat{U}^\dagger_\alpha}{\sqrt{K_\alpha}}\Big)\hat{A}\ket{\varphi_\alpha}+\bra{\varphi_\alpha}\hat{A}\partial_\alpha\Big(\frac{\hat{U}_\alpha}{\sqrt{K_\alpha}}\Big)\ket{\psi_0}\nonumber\\
&=&\bra{\psi_0}\frac{(\partial_\alpha\hat{U}^\dagger_\alpha)\sqrt{K_\alpha}-\frac{\partial_\alpha K_\alpha}{2\sqrt{K_\alpha}}\hat{U}_\alpha^\dagger}{K_\alpha}\hat{A}\ket{\varphi_\alpha}+\bra{\varphi_\alpha}\hat{A}\frac{(\partial_\alpha\hat{U}_\alpha)\sqrt{K_\alpha}-\frac{\partial_\alpha K_\alpha}{2\sqrt{K_\alpha}}\hat{U}_\alpha}{K_\alpha}\ket{\psi_0}\nonumber\\
&=&\bra{\psi_0}\frac{\partial_\alpha\hat{U}_\alpha^\dagger}{\sqrt{K_\alpha}}\hat{A}\ket{\varphi_\alpha}+\bra{\varphi_\alpha}\hat{A}\frac{\partial_\alpha\hat{U}_\alpha}{\sqrt{K_\alpha}}\ket{\psi_0}-\frac{\partial_\alpha K_\alpha}{K_\alpha}\langle\hat{A}\rangle_\alpha\nonumber\\
&=&i[\bra{\varphi_\alpha}(-i)(\hat{U}_\alpha^\dagger)^{-1}(\partial_\alpha\hat{U}_\alpha^\dagger)\hat{A}\ket{\varphi_\alpha}-\bra{\varphi_\alpha}i\hat{A}(\partial_\alpha\hat{U}_\alpha)\hat{U}_\alpha^{-1}\ket{\varphi_\alpha}]-\frac{\partial_\alpha K_\alpha}{K_\alpha}\langle\hat{A}\rangle_\alpha\nonumber\\
&=&i(\langle \hat{h}^{\dagger}\hat{A}\rangle_\alpha-\langle\hat{A} \hat{h}\rangle_\alpha)-\frac{\partial_\alpha K_\alpha}{K_\alpha}\langle\hat{A}\rangle_\alpha,\label{Q}
\end{eqnarray}
where $K_\alpha=\bra{\psi_0}\hat{U}_\alpha^\dagger\hat{U}_\alpha\ket{\psi_0}=\langle\psi_\alpha|\psi_\alpha\rangle$ is the normalized coefficient, then $(\partial_\alpha K_\alpha)/K_\alpha$ can be further written as
\begin{eqnarray}
\frac{\partial_\alpha K_\alpha}{K_\alpha}&=&\frac{\bra{\psi_0}(\partial_\alpha\hat{U}_\alpha^\dagger)\hat{U}_\alpha\ket{\psi_0}+\bra{\psi_0}\hat{U}_\alpha^\dagger(\partial_\alpha\hat{U}_\alpha)\ket{\psi_0}}{K_\alpha}\nonumber\\
&=&\frac{\bra{\psi_\alpha}(\hat{U}_\alpha^\dagger)^{-1}(\partial_\alpha\hat{U}_\alpha^\dagger)\ket{\psi_\alpha}+\bra{\psi_\alpha}(\partial_\alpha\hat{U}_\alpha)\hat{U}_\alpha^{-1}\ket{\psi_\alpha}}{K_\alpha}\nonumber\\
&=&i(\langle \hat{h}^{\dagger}\rangle_\alpha-\langle \hat{h}\rangle_\alpha).
\end{eqnarray}
Substitute this result into Eq. (\ref{Q}), we obtain
\begin{eqnarray}\label{Q2}
|Q|^2&=&|(\langle \hat{h}^{\dagger}\hat{A}\rangle_\alpha-\langle\hat{A} \hat{h}\rangle_\alpha)-(\langle \hat{h}^{\dagger}\rangle_\alpha-\langle \hat{h}\rangle_\alpha)\langle\hat{A}\rangle_\alpha|^2\nonumber\\
&=&-(\langle \hat{h}^{\dagger} \hat{A}\rangle_\alpha-\langle\hat{A}\hat{h}\rangle_\alpha)^2-(\langle \hat{h}^{\dagger}\rangle_\alpha-\langle \hat{h}\rangle_\alpha)^2\langle \hat{A}\rangle_\alpha^2+2(\langle \hat{h}^{\dagger}\rangle_\alpha-\langle \hat{h}\rangle_\alpha)(\langle \hat{h}^{\dagger} \hat{A}\rangle_\alpha-\langle\hat{A}\hat{h}\rangle_\alpha)\langle \hat{A}\rangle_\alpha\nonumber\\
&=&-(\langle \hat{h}^{\dagger} \hat{A}\rangle_\alpha+\langle\hat{A}\hat{h}\rangle_\alpha)^2-(\langle \hat{h}^{\dagger}\rangle_\alpha+\langle \hat{h}\rangle_\alpha)^2\langle \hat{A}\rangle_\alpha^2+2(\langle \hat{h}^{\dagger}\rangle_\alpha-\langle \hat{h}\rangle_\alpha)(\langle \hat{h}^{\dagger} \hat{A}\rangle_\alpha-\langle\hat{A}\hat{h}\rangle_\alpha)\langle \hat{A}\rangle_\alpha\nonumber\\
&&+4\big[\langle \hat{h}^{\dagger} \hat{A}\rangle_\alpha\langle\hat{A}\hat{h}\rangle_\alpha+\langle \hat{h}^{\dagger}\rangle_\alpha\langle \hat{h}\rangle_\alpha\langle \hat{A}\rangle_\alpha^2]\nonumber\\
&=&-(\langle \hat{h}^{\dagger} \hat{A}\rangle_\alpha+\langle\hat{A}\hat{h}\rangle_\alpha)^2-(\langle \hat{h}^{\dagger}\rangle_\alpha+\langle \hat{h}\rangle_\alpha)^2\langle \hat{A}\rangle_\alpha^2+2(\langle \hat{h}^{\dagger}\rangle_\alpha+\langle \hat{h}\rangle_\alpha)(\langle \hat{h}^{\dagger} \hat{A}\rangle_\alpha+\langle\hat{A}\hat{h}\rangle_\alpha)\langle \hat{A}\rangle_\alpha\nonumber\\
&&+4\big[\langle \hat{h}^{\dagger} \hat{A}\rangle_\alpha\langle\hat{A}\hat{h}\rangle_\alpha+\langle \hat{h}^{\dagger}\rangle_\alpha\langle \hat{h}\rangle_\alpha\langle \hat{A}\rangle_\alpha^2-\langle \hat{h}^{\dagger} \hat{A}\rangle_\alpha\langle \hat{h}\rangle_\alpha\langle \hat{A}\rangle_\alpha-\langle\hat{A}\hat{h}\rangle_\alpha\langle \hat{h}^{\dagger}\rangle_\alpha\langle \hat{A}\rangle_\alpha\big].
\end{eqnarray}
Here, we define an operator $\delta\hat{M}=\hat{M}-\langle \hat{M}\rangle_\alpha$, the variance of $\hat{M}$ over $\ket{\varphi_\alpha}$ is defined by
\begin{eqnarray}
(\Delta\hat{M})^2=\bra{\varphi_\alpha}(\hat{M}-\langle M\rangle_\alpha)^\dagger(\hat{M}-\langle M\rangle_\alpha)\ket{\varphi_\alpha}=\bra{\varphi_\alpha}\delta\hat{M}^\dagger\delta\hat{M}\ket{\varphi_\alpha}.
\end{eqnarray}
According to the non-Hermitian uncertainty relationship $(\Delta\hat{A})^2(\Delta\hat{B})^2\geq|\langle\hat{A}^{\dagger}\hat{B}\rangle-\langle\hat{A}^{\dagger}\rangle\langle\hat{B}\rangle|^2$ \cite{ur1,ur2,ur3,ur4,zx}, we have
\begin{eqnarray}\label{un-ra}
(\Delta \hat{h})^2(\Delta\hat{A})^2&\geq&|\langle \hat{h}^{\dagger}\hat{A}\rangle_\alpha-\langle \hat{h}^{\dagger}\rangle\langle \hat{A}\rangle_\alpha|^2\nonumber\\
&=&\langle \hat{h}^{\dagger} \hat{A}\rangle_\alpha\langle\hat{A}\hat{h}\rangle_\alpha+\langle \hat{h}^{\dagger}\rangle_\alpha\langle \hat{h}\rangle_\alpha\langle \hat{A}\rangle_\alpha^2-\langle \hat{h}^{\dagger} \hat{A}\rangle_\alpha\langle \hat{h}\rangle_\alpha\langle \hat{A}\rangle_\alpha-\langle\hat{A}\hat{h}\rangle_\alpha\langle \hat{h}^{\dagger}\rangle_\alpha\langle \hat{A}\rangle_\alpha.
\end{eqnarray}
Obviously, the expression in the last brackets of Eq. (\ref{Q2}) is the same as the result in Eq. (\ref{un-ra}), so we have
\begin{eqnarray}
|Q|^2&\leq&4(\Delta \hat{h})^2(\Delta\hat{A})^2-(\langle \hat{h}^{\dagger} \hat{A}\rangle_\alpha+\langle\hat{A}\hat{h}\rangle_\alpha)^2-(\langle \hat{h}^{\dagger}\rangle_\alpha+\langle \hat{h}\rangle_\alpha)^2\langle \hat{A}\rangle_\alpha^2\nonumber\\
&&+2(\langle \hat{h}^{\dagger}\rangle_\alpha+\langle \hat{h}\rangle_\alpha)(\langle \hat{h}^{\dagger} \hat{A}\rangle_\alpha+\langle\hat{A}\hat{h}\rangle_\alpha)\langle \hat{A}\rangle_\alpha\nonumber\\
&\leq&4(\Delta \hat{h})^2(\Delta\hat{A})^2-\big[(\langle \hat{h}^{\dagger} \hat{A}\rangle_\alpha+\langle\hat{A}\hat{h}\rangle_\alpha)-(\langle \hat{h}^{\dagger}\rangle_\alpha+\langle \hat{h}\rangle_\alpha)\langle \hat{A}\rangle_\alpha\big]^2.
\end{eqnarray}
Define $\ket{f}=\delta \hat{h}\ket{\varphi_\alpha}$ and $\ket{g}=\delta\hat{A}\ket{\varphi_\alpha}$, we have $\bra{f}g\rangle=\langle \hat{h}^{\dagger} \hat{A}\rangle_\alpha-\langle \hat{h}^{\dagger}_\alpha\rangle\langle \hat{A}\rangle_\alpha$. Then, we further obtain the inequality as follow
\begin{eqnarray}
|Q|^2&\leq&4(\Delta \hat{h})^2(\Delta\hat{A})^2-\big[(\langle \hat{h}^{\dagger} \hat{A}\rangle_\alpha-\langle \hat{h}^{\dagger}\rangle_\alpha\langle \hat{A}\rangle_\alpha)+(\langle\hat{A}\hat{h}\rangle_\alpha-\langle \hat{h}\rangle_\alpha\langle \hat{A}\rangle_\alpha)\big]^2\nonumber\\
&=&4(\Delta \hat{h})^2(\Delta\hat{A})^2-(\bra{f}g\rangle+\bra{g}f\rangle)\nonumber\\
&\leq&4(\Delta \hat{h})^2(\Delta\hat{A})^2.
\end{eqnarray}
According to Cauchy-Schwarz inequality, the first inequality is saturated when $\ket{f}$ and $\ket{g}$ is linearly related, i.e., $\ket{f}=c\ket{g}$. And the second inequality is further saturated if $c$ is an imaginary number. Therefore, we could derive the inequality
\begin{eqnarray}
(\Delta\alpha)^2=\frac{(\Delta \hat{A})^2}{n|\partial_\alpha \langle \hat{A}\rangle_\alpha|^2}=\frac{(\Delta\hat{A})^2}{n|Q|^2}\geq\frac{(\Delta\hat{A})}{4n(\Delta \hat{h})^2(\Delta\hat{A})^2}=\frac{1}{n\mathcal{F}_\alpha},
\end{eqnarray}
it is exactly the QCRB, as we discussed above, it is saturated when
\begin{eqnarray}
\ket{f}=ic\ket{g},\label{optmeas}
\end{eqnarray}
where $c$ is a real number. \hfill $\square$

\subsection*{\normalsize D. The reason for normalization}
\begin{figure*}[h]
\centering
\includegraphics[width=0.9\textwidth]{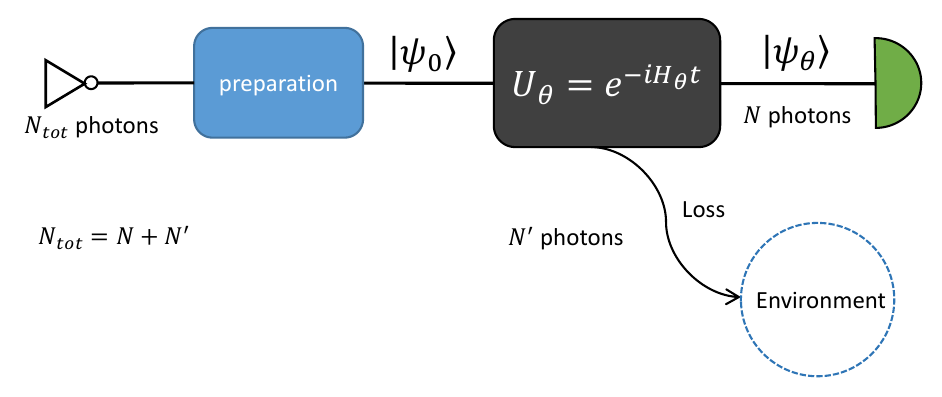}
\caption{Simple diagram of quantum estimation in non-Hermitian system.}\label{normal}
\end{figure*}

It is well-known that the probability may not conserve in non-Hermitian systems evolution. However, for measurement process, the probabilities of measurement outcomes always sum up to 1, even if there are losses or gains during evolution process, and the final state is normalized by dividing its trace. As shown in Fig. \ref{normal}, we use the polarization of photons to encode states, where horizontally polarized state $|H\rangle$ is encoded as $|0\rangle$, and vertically polarized state $|V\rangle$ is encoded as $|1\rangle$, the photons are prepared as initial state $\rho_0=|\psi_0\rangle\langle\psi_0|$ and then evolves in the non-Hermitian quantum system, which will lead to the decay of the total number of photons $N_{tot}$, i.e., $N=N_{tot}\mathrm{Tr}\{\rho_\theta\}<N_{tot}$. However, the probability is still normalized when we measure the final state $\rho_\theta=U_\theta\rho_0U^{\dagger}_\theta$. For instance, we measure the final state with the set of projection operation $\Pi=\sum_{i}|i\rangle\langle i|$, the probabilities of two measurement outcomes respectively are $P_0=N_H/N=(N_{tot}\mathrm{Tr}\{\rho_\theta|0\rangle\langle 0|\})/(N_{tot}\mathrm{Tr}\{\rho_\theta\})$ and $P_1=N_V/N=(N_{tot}\mathrm{Tr}\{\rho_\theta|1\rangle\langle 1|\})/(N_{tot}\mathrm{Tr}\{\rho_\theta\})$, the total probability is $P=P_0+P_1=1$, where $N_H$ and $N_V$ respectively are the number of horizontally polarized photons and vertically polarized photons. For measurement, the final state is actually
\begin{eqnarray}
\widetilde{\rho}_\theta=\rho_\theta/\mathrm{Tr}\{\rho_\theta\}.
\end{eqnarray}
Therefore, we normalize the final state by dividing its trace, for pure states, the normalized final state is
\begin{eqnarray}
|\varphi_\theta\rangle=\frac{|\psi_\theta\rangle}{\sqrt{\langle\psi_\theta|\psi_\theta\rangle}}.
\end{eqnarray}

\newpage

\section*{\large \uppercase\expandafter{\romannumeral2}.  EXPERIMENT}
\begin{figure*}[h]
\centering
\includegraphics[width=0.35\textwidth]{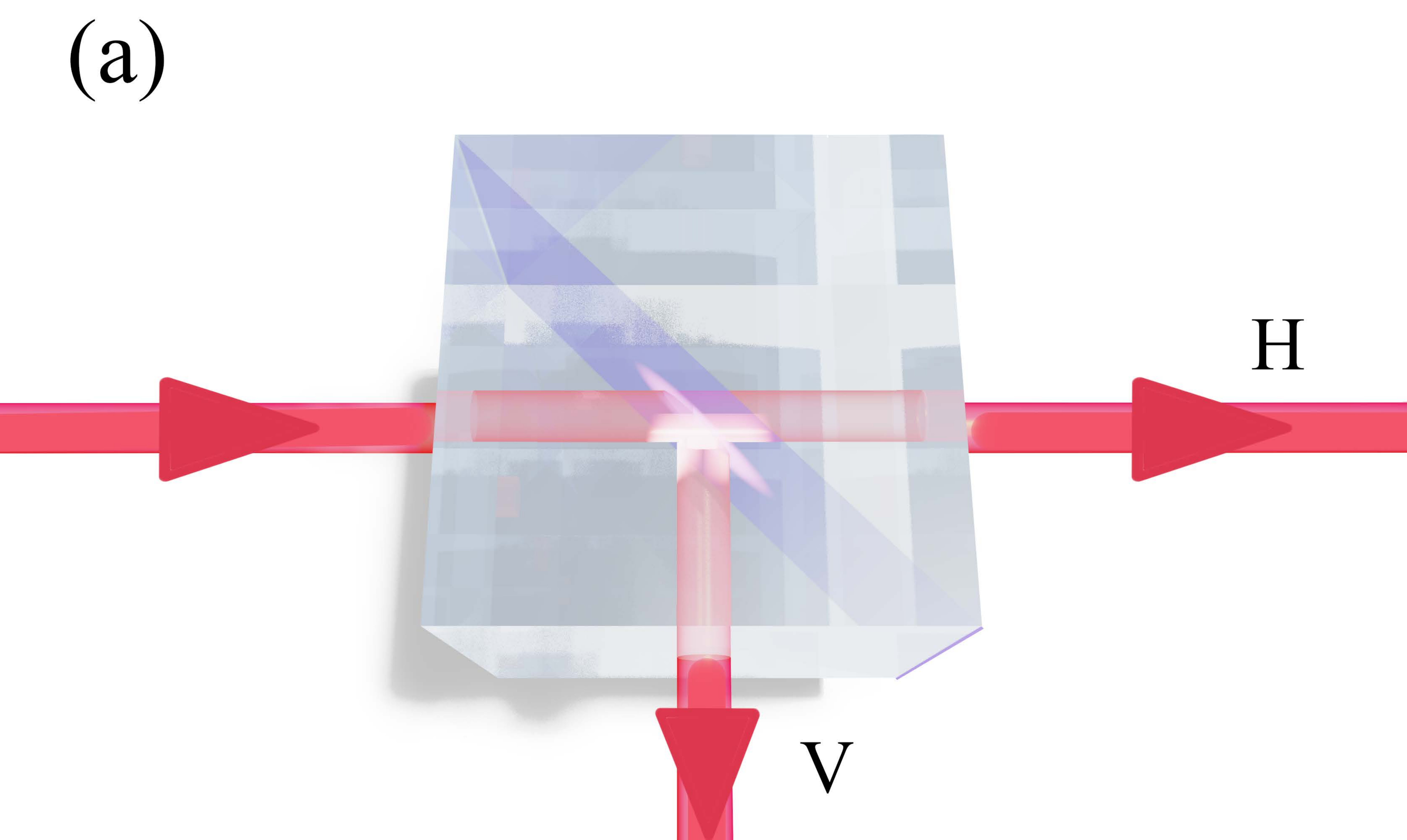}\hspace{30mm}
\includegraphics[width=0.35\textwidth]{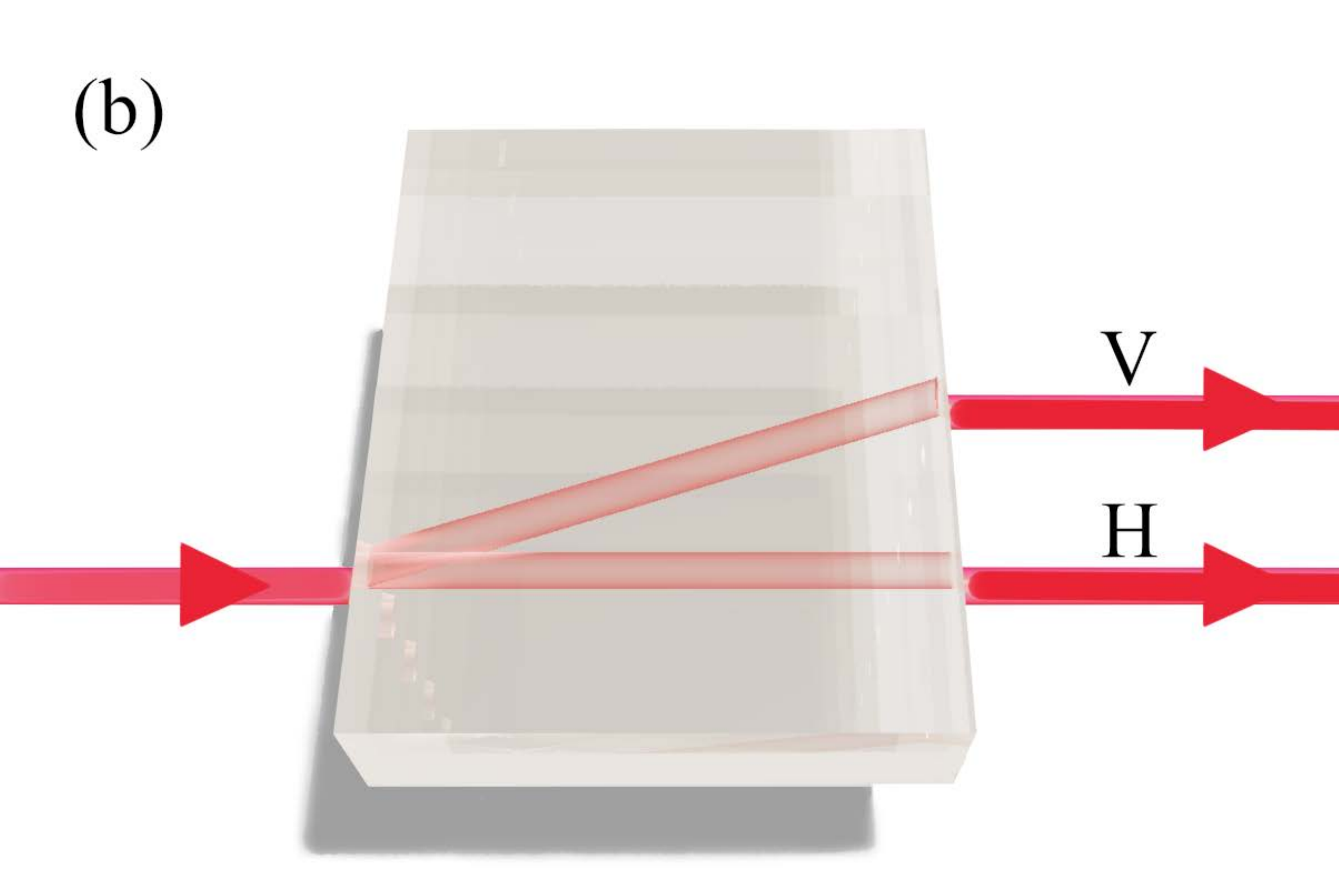}
\caption{Schematic illustration of PBS and BD.}\label{optic}
\end{figure*}

Before we introduce our experimental setup, we first illustrate the optics elements in our experiment.
\begin{itemize}
  \item [1)]
  The Jones matrix of a half-wave plate (HWP) and quarter-wave plate (QWP) used in our experiment are given by:
  \begin{eqnarray}
  U_{HWP}=\begin{pmatrix} \cos2\theta & \sin2\theta \\ \sin2\theta & -\cos2\theta  \end{pmatrix}, \quad U_{QWP}=\begin{pmatrix} \cos^2\theta+i\sin^2\theta & (1-i)\cos\theta\sin\theta \\ (1-i)\cos\theta\sin\theta & i\cos^2\theta+\sin^2\theta  \end{pmatrix},\nonumber\\
  \end{eqnarray}
where $\theta$ is the angle between the light polarization direction and the fast axis of the wave plate.
  \item [2)]
  The polarization beam splitter (PBS) in our experiment transmits the horizontal polarized photons and reflect the vertical polarized photons, as shown in Fig. \ref{optic}(a).
  \item [3)]
  The beam displacer (BD) in our experiment transmits the horizontal polarized photons, but separates the vertical photons into new path which is about 4mm from the original path, as shown in Fig. \ref{optic}(b).
\end{itemize}

\begin{figure*}[t]
\centering
\includegraphics[width=0.85\textwidth]{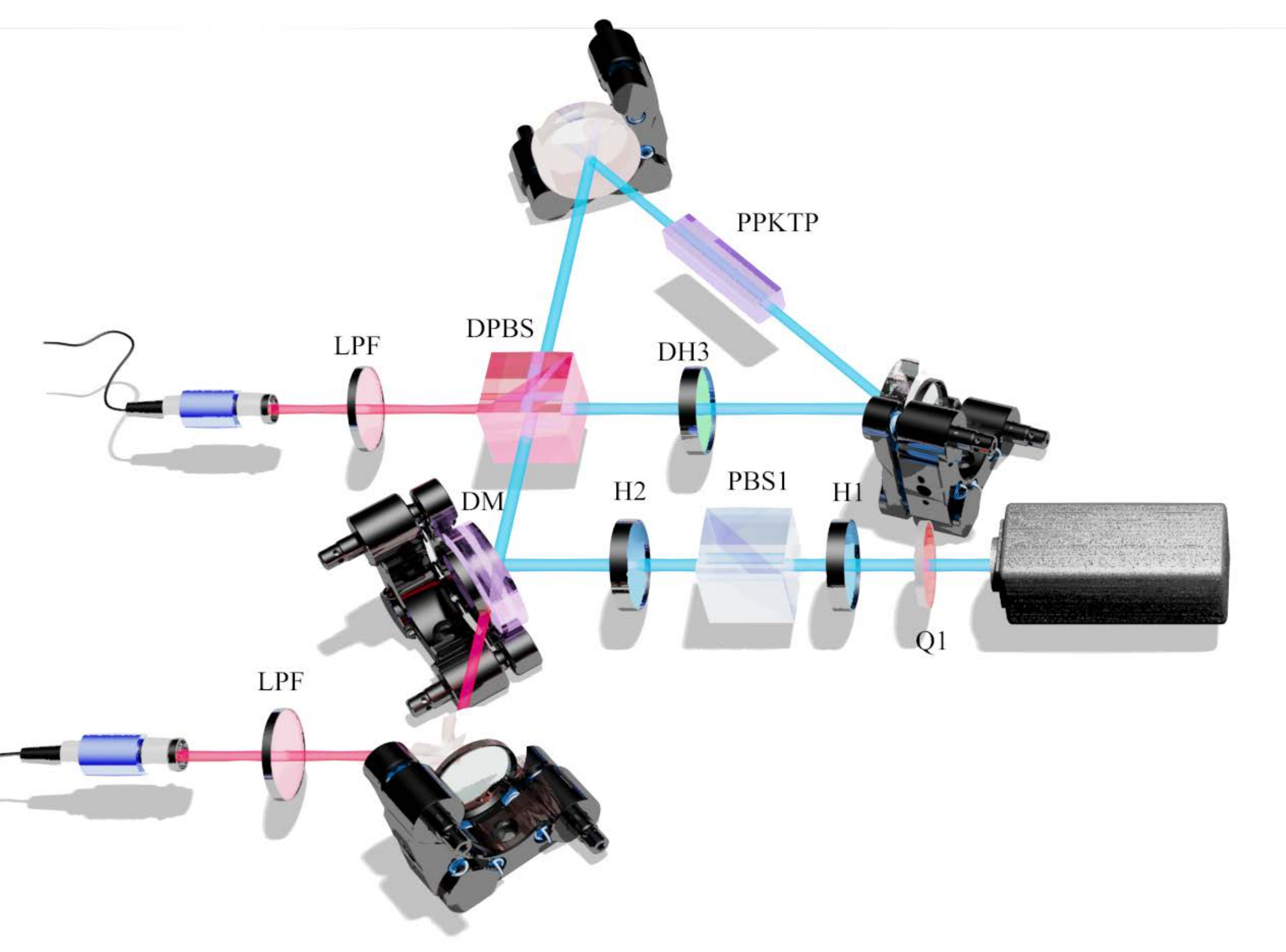}
\caption{Schematic illustration of the photon-pair source.}\label{fig1}
\end{figure*}

\subsection*{\normalsize A. Photon-pair source}
The central wavelength of pump laser in our experiment is $405$ nm. The energy of pump laser is adjust by QWP1, HWP1 and PBS1, meanwhile the state is purified into $\ket{H}$. If the initial state is prepared as $(\ket{H}+\ket{V})/\sqrt{2}$, this experimental set could generate entangled photon pair $(\ket{HV}+\ket{VH})/\sqrt{2}$. In our experiment, only the heralded single photon source is needed, so we prepare the initial state as $\ket{H}$. A dichroic mirror (DM) reflect pump light into the triangle sagnac interferometer, and then PPKTP crystal is clockwise pumped. After type-II phase-matched spontaneous parametric down-conversion process, one 405 nm photon splits into two 810 nm photons ($\ket{H}\rightarrow\ket{H}_1\ket{V}_2$).  The DPBS and two mirrors are affective for both 405 nm and 810 nm photons, but DM does not reflect 810 nm photons. So these two 810nm photons are separated into two paths, then filtered by long-wave path filter (LPF) and collected into single-mode fibers. In our experiment, the power of pumped light is 2 mw, and we could obtain 80000 coincidences per second.

\begin{figure}[h]
\centering
\includegraphics[ width=0.9\textwidth]{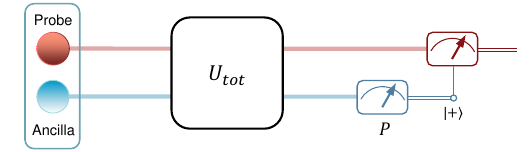}
\caption{Post-selected scheme for the non-unitary evolution. The operator $U_{tot}$ is an evolution with loss process, we effectively obtain the evolution $\hat{U}'_{PT}=F\hat{U}_{PT}$ of the $PT$-symmetric Hamiltonian $\hat{H}_{PT}$ for probe qubit after post-selection.  }\label{circuit}
\end{figure}

\begin{figure*}[b]
\centering
    \includegraphics[width=1\textwidth]{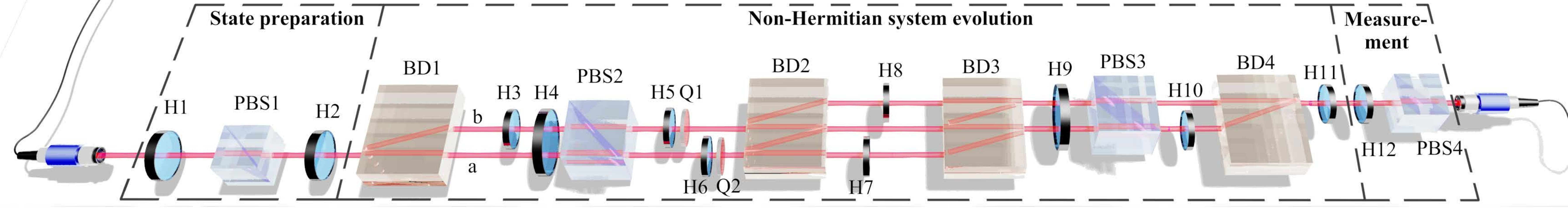}
\caption{Schematic illustration of the non-Hermitian system evolution.}\label{fig2}
\end{figure*}
\subsection*{\normalsize B. Theoretical framework of $\hat{U}_{PT}$ evolution}
The non-unitary $\mathcal{PT}$-symmetric evolution $\hat{U}_{PT}$ consists of the probe qubit ($\ket{H,V}$) and ancilla qubit ($\ket{a,b}$), by performing a projective operator on ancilla qubit, we could effectively construct a non-unitary $\mathcal{PT}$-symmetric evolution $\hat{U}'_{PT}=F\hat{U}_{PT}$ \cite{spn1}, as shown in Fig.\;\ref{circuit}. The non-unitary $\mathcal{PT}$-symmetric evolution $\hat{U}_{PT}$ can be written as $\hat{U}_{PT}=|\psi_H\rangle\langle H|+|\psi_V\rangle\langle V|$ ($\ket{\psi_H}=\hat{U}_{PT}\ket{H}$,  $\ket{\psi_V}=\hat{U}_{PT}\ket{V}$). The operator $\hat{U}_{tot}$ is unitary, but in Fig.\;\ref{circuit}, we do not concentrate on the photons (a new qubit) losses at H3, H4 and PBS2, the left two-qubit evolution $\hat{U}$ can be expressed as follow,
\begin{eqnarray}
\hat{U}&=&\gamma(|\psi_H\rangle\langle H|\otimes|a\rangle\langle a|+|\psi_V\rangle\langle V|\otimes|b\rangle\langle a|+|\psi_H^\perp\rangle\langle V|\otimes|a\rangle\langle b|+|\psi_V^\perp\rangle\langle H|\otimes|b\rangle\langle b|),\nonumber\\
&=&p|\varphi_H\rangle\langle H|\otimes|a\rangle\langle a|+q|\varphi_V\rangle\langle V|\otimes|b\rangle\langle a|+p|\varphi_H^\perp\rangle\langle V|\otimes|a\rangle\langle b|+q|\varphi_V^\perp\rangle\langle H|\otimes|b\rangle\langle b|,\nonumber\\
\end{eqnarray}
where $p=\sin2(\phi_1-\phi_2)$ and $q=\cos2\phi_2$ are controlled by H3 ($\phi_1$) and H4 ($\phi_2$), $p^2/q^2=\langle\psi_H|\psi_H\rangle/\langle\psi_V|\psi_V\rangle$, $\ket{\varphi_H}=\ket{\psi_H}/\sqrt{\langle\psi_H|\psi_H\rangle}$, $\ket{\varphi_V}=\ket{\psi_V}/\sqrt{\langle\psi_V|\psi_V\rangle}$ and $\gamma=p/\sqrt{\langle\psi_H|\psi_H\rangle}=q/\sqrt{\langle\psi_V|\psi_V\rangle}$, if $p=q=1$, $\hat{U}=\hat{U}_{tot}$ is an unitary evolution. As shown in Fig.\;\ref{fig2}, $p$ and $q$ are controlled by H3, H4 and PBS2, $\ket{\varphi_H}$ and $\ket{\varphi_V}$ are prepared by H5, Q1, H6 and Q1 respectively, $\ket{\varphi_H^\perp}$ and $\ket{\varphi_V^\perp}$ are the corresponding orthogonal states. After preparing $\ket{\varphi_H}\ket{a}$ and $\ket{\varphi_V}\ket{b}$, BD2 and BD3 combine their horizontal and vertical components into two paths respectively. Then the projective operator is realized by PBS3 and H9 set at $22.5^\circ$, it can be expressed as
\begin{eqnarray}
\hat{P}=I\otimes\frac{\ket{a}+\ket{b}}{\sqrt{2}}\frac{\bra{a}+\bra{b}}{\sqrt{2}},
\end{eqnarray}
where $I$ is identity operator. The evolution after performing projective operator is
\begin{eqnarray}
\hat{U}_{PT}'=\hat{P}\hat{U}=\frac{p|\varphi_H\rangle\langle H|+q|\varphi_V\rangle\langle V|}{\sqrt{2}}\otimes\frac{\ket{a}+\ket{b}}{\sqrt{2}}\bra{a}.
\end{eqnarray}
In our experiment, the initial state of ancilla qubit is always $\ket{a}$, then, for an arbitrary initial state $\ket{\psi_0}=\cos2\phi\ket{H}+\sin2\phi\ket{V}$ of probe qubit, the final state after evolution is $(p\cos2\phi\ket{\varphi_H}+q\sin2\phi\ket{\varphi_V})/\sqrt{2}$. Thus,  for probe qubit the evolution we construct is proportional to the theoretical non-Hermitian system evolution,
\begin{eqnarray}
\hat{U}_{PT}'=\frac{p|\varphi_H\rangle\langle H|+q|\varphi_V\rangle\langle V|}{\sqrt{2}}=\frac{\gamma}{\sqrt{2}}(|\psi_H\rangle\langle H|+|\psi_V\rangle\langle V|)=F\hat{U}_{PT}.
\end{eqnarray}
where $F=\gamma/\sqrt{2}$ is a scalar function.

\subsection*{\normalsize C. Optimal measurements}
In this section, we prove that $\hat{A}=|0\rangle\langle0|$ is optimal measurement for both multiplicative and non-multiplicative Hamiltonians when the probe state is $\ket{0}$. In the case of estimate $s$, according to the condition for optimal measurements.
\begin{eqnarray}
\ket{f}=i\frac{t\cos\alpha\cos(\alpha-2st\cos\alpha)\sin\alpha}{\cos^2(a-2st\cos\alpha)+\sin^2(st\cos\alpha)}\ket{g},
\end{eqnarray}
so the measurement $\hat{A}$ is indeed the optimal measurement when estimate $s$.

In the case of estimate $\alpha$, we further experimentally verify the result. We first give the theoretical analysis.t he probe state is prepared as an arbitrary linear polarization pure state $\ket{\psi_0}=\cos2\phi\ket{0}+\sin2\phi\ket{1}$, we change the probe state from $\ket{0}$ to $\ket{1}$ ($\phi=0^\circ\thicksim45^\circ$) and perform $\hat{A}=|0\rangle\langle0|$ as measurement. According to Eq\;(\ref{optmeas}), we have
\small{
\begin{eqnarray}
&&\ket{f}\nonumber\\
&=&\frac{2\cos\alpha[st\sin\alpha(\cos4\alpha-i\sin4\phi)+\sin^2(st\cos\alpha)]+(i\sin4\phi-\cos4\phi\sin\alpha)\sin(2st\cos\alpha)}{2[\cos2\phi\cos(\alpha-st\cos\alpha)-i\sin2\phi\sin(st\cos\alpha)][\cos(\alpha+st\cos\alpha)\sin2\phi-i\cos2\phi\sin(st\cos\alpha)]}\ket{g},\nonumber\\
\end{eqnarray}}
we can see that the condition for optimal measurement is satisfied only if $\phi=k\pi/4$, i.e., the probe state is $\ket{0}$ or $\ket{1}$.

\begin{figure}[b]
\centering
\includegraphics[ width=0.8\textwidth]{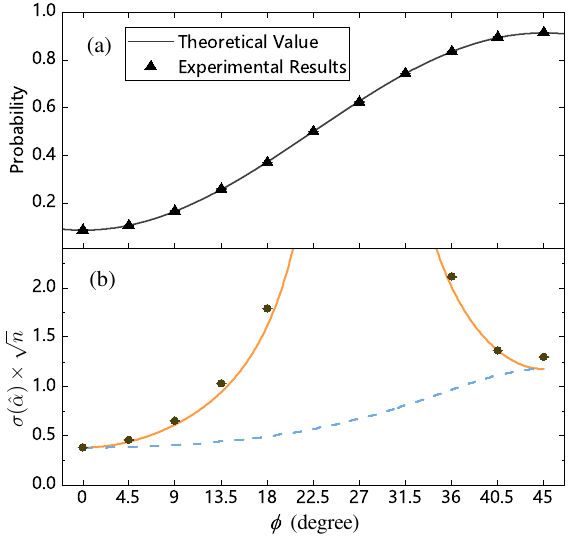}
\caption{Estimation precision. (a) The probabilities of measurement outcomes for different probe states after evolution. We set the measurement as $\hat{A}=|0\rangle\langle0|$, the probe states $\ket{\psi_0}=\cos2\phi\ket{0}+\sin2\phi\ket{1}$ is changed from $\phi=0^\circ$ to $\phi=45^\circ$. (b) The standard deviation. The experimental standard deviation (black dots) approximately matches well the theoretical estimation precision (orange solid lines) calculated by the error-propagation formula. The ultimate precision described by QFI (blue dashed lines) is achieved when the probe state is $\ket{0}$ or $\ket{1}$.}\label{result}
\end{figure}

In our experiment, we set that $s=1$, $\alpha=\pi/10$ and $t=\pi/[2s\cos(\pi/10)]$. To get the statistic of the estimation, we make $5000$ maximum likelihood estimates and get the distribution of estimators $\hat{\alpha}$, the experimental results is shown in Fig.\;\ref{result}. Due to the fluctuation of the data, the estimator $\hat{\alpha}$ cannot be calculated by maximum likelihood estimation when $\phi=22.5^\circ$, $\phi=27^\circ$ and $\phi=31.5^\circ$, so the corresponding deviations are not plotted. As shown in Fig.\;\ref{result}(b), the standard deviation of estimator reaches QCRB when the probe state is $\ket{0}$ and $\ket{1}$, it is consistent with our theoretical analysis. We also give the exact data of Fig.\;\ref{result}, as shown in Table.\;\ref{tabopt} and \ref{taboptestmt}.

\begin{table*}[t]
  \centering
  \normalsize
  \setlength{\tabcolsep}{10pt}
  \renewcommand\arraystretch{1.3}
  \begin{tabular}{ccc}
  \hline
  \toprule
    \large{$\phi$}  & Experiment $p_0$ & Theoretical $p_0$  \\
  \hline
  $0^\circ$ & $0.087393\pm0.000051$ & 0.0872   \\

  $4.5^\circ$ & $0.106596\pm0.000053$ & 0.1074  \\

  $9^\circ$ & $0.167888\pm0.000065$ & 0.1660  \\

  $13.5^\circ$ & $0.258250\pm0.000075$ & 0.2573  \\

  $18^\circ$ & $0.370123\pm0.000087$ & 0.3724  \\

  $22.5^\circ$ & $0.499996\pm0.000089$ & 0.5000  \\

  $27^\circ$ & $0.622651\pm0.000086$ & 0.6276 \\

  $31.5^\circ$ & $0.742903\pm0.000078$ & 0.7427 \\

  $36^\circ$ & $0.833638\pm0.000064$ & 0.8340  \\

  $40.5^\circ$ & $0.896258\pm0.000050$ & 0.8926   \\

  $45^\circ$ & $0.913169\pm0.000045$ & 0.9128  \\
  \bottomrule
  \hline
  \end{tabular}
  \caption{The measurement results of $\hat{\alpha}$ for probe states.}\label{tabopt}
\end{table*}

\begin{table*}[t]
   \begin{tabular}{cccc}
   \toprule
   \hline
   \large{$\phi$}   & Experiment \large{$\sigma(\hat{\alpha})$} & Theoretical \large{$\sigma(\hat{\alpha})$} & \large{$1/\sqrt{\mathcal{F}_\alpha}$}  \\
  \hline
 $0^\circ$  & $0.3823\pm0.0001$ & 0.3813 &  0.3813  \\

  $4.5^\circ$ & $0.4556\pm0.0002$ & 0.4391 & 0.3875 \\

  $9^\circ$  & $0.6520\pm0.0002$ & 0.6128 & 0.4068 \\

  $13.5^\circ$ & $1.0282\pm0.0004$ & 0.9425 & 0.4409 \\

  $18^\circ$ & $1.7911\pm0.0006$ & 1.6165 & 0.4934 \\

  $22.5^\circ$  & ------ & 3.6517 & 0.5689 \\

  $27^\circ$  & ------ & 133.1113 & 0.6732  \\

  $31.5^\circ$  & ------ & 3.9067 & 0.8094 \\

  $36^\circ$ & $2.1140\pm0.0008$ & 2.0078 & 0.9690 \\

  $40.5^\circ$  & $1.3645\pm0.0005$ & 1.3661 & 1.1142  \\

  $45^\circ$  & $1.2968\pm0.0005$ & 1.1763 &  1.1763 \\
  \bottomrule
  \hline
  \end{tabular}
  \caption{The standard deviations of $\hat{\alpha}$ for probe states.}\label{taboptestmt}
\end{table*}

\begin{figure}[t]
\centering
\includegraphics[ width=1.0\textwidth]{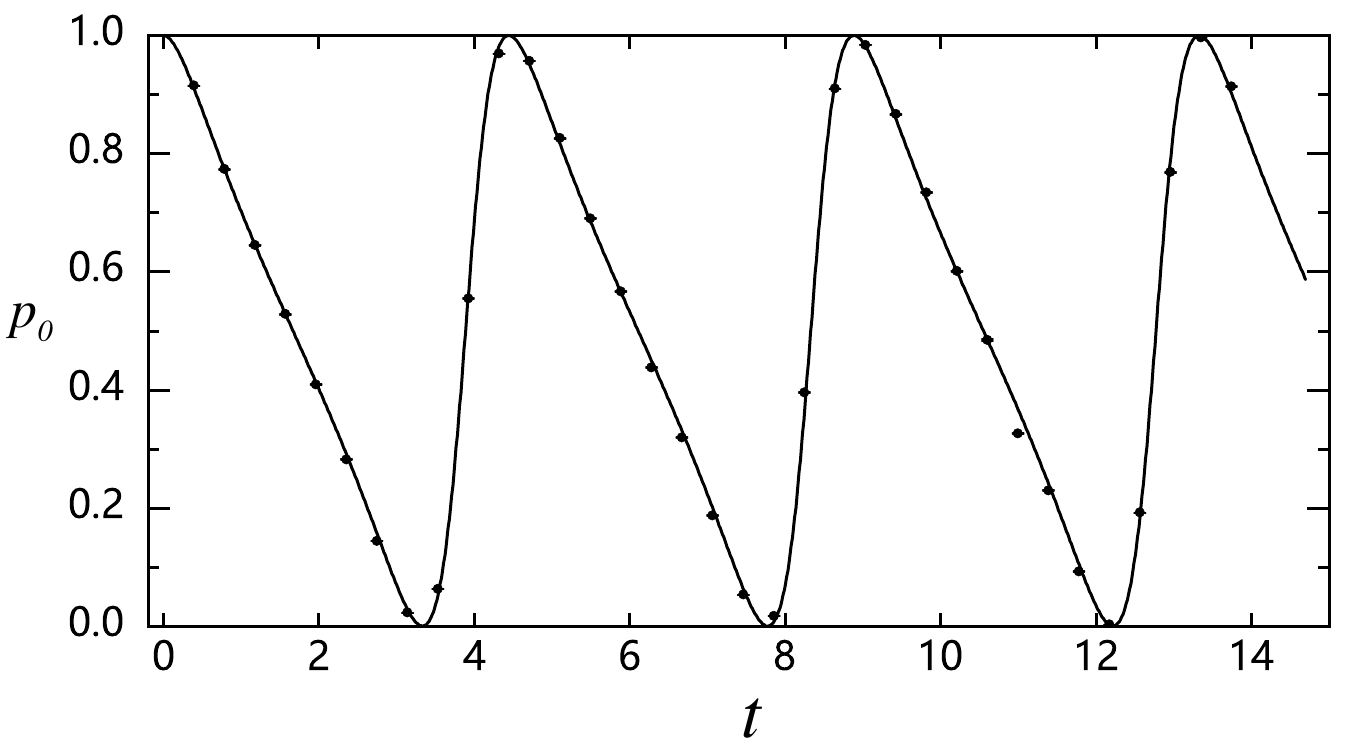}
\caption{The probabilities of measurement outcomes for varying t. The black dots represent the experimentally measured data of $P_0$ for varying $t$. And we set $s=1$, $\alpha=\pi/4$, the measurement performed is $\hat{A}=|0\rangle\langle0|$ and the probe state is $\ket{\psi_0}=\ket{0}$. The black solid line is the theoretical value of $p_0=\bra{\varphi}\hat{A}\ket{\varphi}$, the data points match well with the theoretical curve.}\label{probability}
\end{figure}

\subsection*{\normalsize D. Maximum likelihood estimation}
When we estimate the parameter with different probe states, we apply maximum likelihood estimation to get the estimator $\hat{\alpha}$. For two-level systems, the distribution of measurement outcomes obeys binomial distribution, for $n$ independent measurements, the likelihood function is
\begin{eqnarray}
L(n,x|\alpha)=\frac{n!}{(n-x)!x!}p(\alpha)^x[1-p(\alpha)]^{(n-x)}.
\end{eqnarray}
This expression represents the probability that there are $x$ measurement outcomes $\ket{0}$ in $n$ measurements, and $p(\alpha)=|\langle0|\varphi_\alpha\rangle|^2$ is the probability of obtaining $\ket{0}$. The logarithmic of the likelihood function is
\begin{eqnarray}
\ln[L(n,x|\alpha)]=\ln[\frac{n!}{(n-x)!x!}]+x\ln[p(\alpha)]+(n-x)\ln[1-p(\alpha)].
\end{eqnarray}
Then we solve the differential equation
\begin{eqnarray}
0=\frac{d\ln[L(n,x|\alpha)]}{d\alpha}=x\frac{d\ln[p(\alpha)]}{d\alpha}+(n-x)\frac{d\ln[1-p(\alpha)]}{d\alpha}.
\end{eqnarray}
The solution of the equation is the estimator $\hat{\alpha}=\hat{\alpha}(n,x)$, which is the function of $n$ and $x$. By substituting the experimental results into $\hat{\alpha}(n,x)$, we obtain the estimate of the parameter $\alpha$. We further obtain the distribution of estimations by repeating this process.

\subsection*{\normalsize E. Experimental data of reaching Heisenberg scaling}
In the main text, we show the standard deviation of estimation results and a part of the distribution of estimators, here we present more detailed results. As shown in Table.\;\ref{tabs} and Table.\;\ref{tabs}, we give the exact experimental and theoretical results for different time points. In Table.\;\ref{tabs} and Table.\;\ref{tabalpha}, we can see that the average of estimators $\hat{s}$ and $\hat{\alpha}$ is slightly different from the theoretical value $s=1$ and $\alpha=\pi/4$, because of the error of the non-unitary $\mathcal{PT}$-symmetric evolution $\hat{U}'_{PT}$ we constructed. As shown in Fig.\;\ref{probability} and Table.\;\ref{tabalp0t}, we can see that although the errors of the probabilities we measured are quite small, there are still errors between the average of estimators and theoretical value, especially for the probe states that lead to worse estimation precision.

\begin{table*}[h]
  \centering
  \normalsize
  \setlength{\tabcolsep}{7pt}
  \renewcommand\arraystretch{1.5}
  \begin{tabular}{ccccc}\toprule\hline
  $t$ & Experiment 1/\large{$\sigma(\hat{s})$} & \large{$\sqrt{\mathcal{F}_s(t)}$} & $E[\hat{s}]$ & error(s) \\
  \hline
  $\pi/8$  & $0.4612\pm0.0146$  & 0.4682 & $0.9644\pm0.0018$ & 3.5640\% \\

  $2\pi/8$  & $0.6449\pm0.0204$  & 0.6406 & $0.9968\pm0.0012$ & 0.3220\% \\

  $3\pi/8$  & $0.7849\pm0.0248$  & 0.7624 & $1.0010\pm0.0010$ & -0.0997\%\\

  $4\pi/8$  & $0.9083\pm0.0287$  & 0.9236 & $0.9982\pm0.0009$ & 0.1785\% \\

  $5\pi/8$  & $1.1755\pm0.0372$  & 1.1933 & $1.0057\pm0.0007$ & -0.5673\% \\

  $6\pi/8$  & $1.7708\pm0.0560$ & 1.6875 & $1.0104\pm0.0005$ & -1.0438\% \\

  $7\pi/8$  & $2.7532\pm0.0871$  & 2.6743 & $1.0123\pm0.0003$ & -1.2325\% \\

  $8\pi/8$  & $4.9603\pm0.1569$  & 4.8245 & $1.0064\pm0.0002$ & -0.6429\\

  $9\pi/8$  & $9.1263\pm0.2887$  & 9.3179 & $1.0043\pm0.0001$ & -0.4260\% \\

  $10\pi/8$  & $13.1820\pm0.4171$  & 13.3574 & $0.9981\pm0.0001$ & 0.1869\%\\
  \bottomrule
  \hline

  \end{tabular}
  \caption{The standard deviations of $\hat{\alpha}$ for different times.}\label{tabs}
\end{table*}

\begin{table*}[h]
  \centering
  \normalsize
  \setlength{\tabcolsep}{7pt}
  \renewcommand\arraystretch{1.5}
  \begin{tabular}{ccccc}\toprule\hline
  $t$ & Experiment 1/\large{$\sigma(\hat{\alpha})$} & \large{$\sqrt{\mathcal{F}_\alpha(t)}$} & $E[\hat{\alpha}]$ & error($\alpha$) \\
  \hline

  $2\pi/8$  & $0.4310\pm0.0136$  & $0.4445$ & $0.7934\pm0.0019$ & 1.0240\% \\

  $3\pi/8$  & $0.8284\pm0.0262$  & $0.8080$ & $0.7855\pm0.0010$ & 0.0170\% \\

  $4\pi/8$  & $1.2346\pm0.0391$  &$ 1.2604$ & $0.7872\pm0.0007$ & 0.2267\% \\

  $5\pi/8$  & $1.8511\pm0.0586$  & $1.8711$ & $0.7819\pm0.0004$ & 0.4403\% \\

  $6\pi/8$  & $2.9330\pm0.0928$ & $2.7872$ & $0.7791\pm0.0003$ & 0.8009\% \\

  $7\pi/8$  & $4.4336\pm0.1403$  & $4.3343$ & $0.7778\pm0.0002$ & 0.9730\% \\

  $8\pi/8$  & $7.3842\pm0.2937$  & $7.2462$ & $0.7811\pm0.0001$ & 0.5491\% \\

  $9\pi/8$  & $12.0692\pm0.3819$  & $12.4451$ & $0.7822\pm0.0001$ & 0.4091\% \\

  $10\pi/8$  & $15.4574\pm0.4891$  & $15.5728$ & $0.7870\pm0.0001$ & 0.2029\% \\
  \bottomrule
  \hline

  \end{tabular}
  \caption{The standard deviations of $\hat{s}$ for different times.}\label{tabalpha}
\end{table*}

\begin{table*}[h]
  \centering
  \normalsize
  \setlength{\tabcolsep}{7pt}
  \renewcommand\arraystretch{1.5}
  \begin{tabular}{ccc}\toprule\hline
  $t$ &  Experiment $p_0$ & Theoretical $p_0$ \\
  \hline

  $ \pi/8$  & $ 0.9151\pm0.0005$  & $0.9104$  \\

  $ 2\pi/8$  & $ 0.7742\pm0.0007$  & $0.7733$  \\

  $ 3\pi/8$  & $ 0.6453\pm0.0008$  &$ 0.6457$ \\

  $ 4\pi/8$  & $ 0.5288\pm0.0009$  & $ 0.5279$ \\

  $ 5\pi/8$  & $ 0.4089\pm0.0009$ & $ 0.4123$ \\

  $ 6\pi/8$  & $ 0.2822\pm0.0007$  & $ 0.2903$ \\

  $ 7\pi/8$  & $ 0.1443\pm0.0006$  & $ 0.1563$ \\

  $ 8\pi/8$  & $ 0.0229\pm0.0003$  & $ 0.0277$ \\

  $ 9\pi/8$  & $ 0.0630\pm0.0004$  & $ 0.0535$ \\

  $ 10\pi/8$  & $ 0.5547\pm0.0008$  & $ 0.5671$ \\
  \bottomrule
  \hline

  \end{tabular}
   \caption{The measurement results for different evolution times.}\label{tabalp0t}
\end{table*}

\clearpage

%\begin{figure*}[t]\label{fig:neigenvalue}
%\centering
%\includegraphics[width=0.8\textwidth]{new_eigenvalue.pdf}
%\R{\caption{The evolution of the modulus of difference $|\lambda_+-\lambda_-|$ as function of time $t$.}}
%\end{figure*}

\subsection*{\normalsize F. Non-Hermitian Hamiltonian without $\mathcal{PT}$ or anti-$\mathcal{PT}$ symmetry}
\begin{figure*}[t]
\centering
\includegraphics[width=0.8\textwidth]{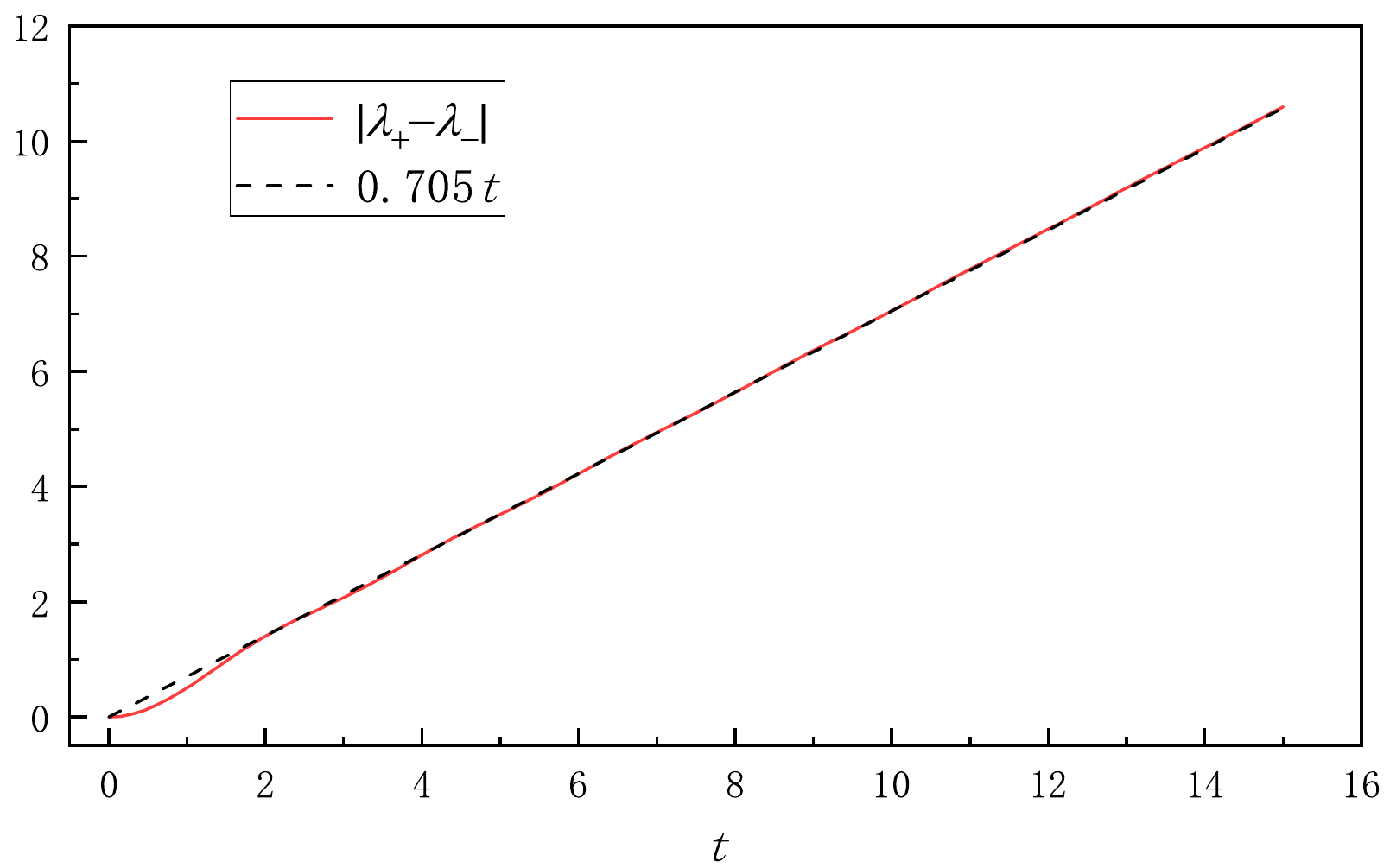}
\caption{The evolution of the modulus of difference $|\lambda_+-\lambda_-|$ as function of time $t$.}\label{fig:neigenvalue}
\end{figure*}

%There is no special assumption on Hamiltonians when derive our theory. 
To demonstrate the generality of our theory, we consider a Hamiltonian without any special symmetries given by
\begin{eqnarray}
\hat{H}_\kappa=\left(\begin{array}{cc} 0 &\quad \kappa \\ 1 & \quad 0 \end{array}\right),
\end{eqnarray}
where $\kappa$ is the unknown real parameter and $\kappa\neq1$. It can be seen that $\hat{H}_\kappa$ is neither $\mathcal{PT}$-symmetric nor anti $\mathcal{PT}$-symmetric. The corresponding evolution operator
$\hat{U}_\kappa$ can be deduced from the Hamiltonian $\hat{H}_\kappa$ as
\begin{eqnarray}
\hat{U}_\kappa(t)=\left(\begin{array}{cc} \cos(t\sqrt{\kappa}) &\quad -i\sqrt{\kappa}\sin(t\sqrt{\kappa}) \\ \frac{-i}{\sqrt{\kappa}}\sin(t\sqrt{\kappa}) & \quad
\cos(t\sqrt{\kappa}) \end{array}\right).
\end{eqnarray}
According to Eq.\;(\ref{h}), we can obtain the generator as
\begin{eqnarray}
\hat{h}_\kappa(t)=\frac{1}{4\kappa\sqrt{\kappa}}\left(\begin{array}{cc} i2\sqrt{\kappa}\sin^2(t\sqrt{\kappa}) &\quad 2t\kappa\sqrt{\kappa}+\kappa\sin(2t\sqrt{\kappa}) \\
2t\sqrt{\kappa}-\sin(2t\sqrt{\kappa}) & -i2\sqrt{\kappa}\sin^2(t\sqrt{\kappa}) \end{array}\right).
\end{eqnarray}
The eigenvalues of $\hat{h}_\kappa(t)$ are $\lambda_\pm=\pm\sqrt{[-1+2\kappa t^2+\cos(2t\sqrt{\kappa})]/(8\kappa)}$. As shown in Fig.\ref{fig:neigenvalue}, the growth of the modulus of the difference,
$|\Delta\lambda|=|\lambda_+-\lambda_-|$, reaches the scale of $t$, indicating that the growth of the quantum Fisher information scales as $t^2$. The initial probe state is prepared as $|\psi_0\rangle=|0\rangle$. By utilizing Eq.\;(\ref{QFI}), we can calculate the QFI as
\begin{eqnarray}
\mathcal{F}_\kappa(t)=\frac{[-2t\sqrt{\kappa}+\sin(2t\sqrt{\kappa})]^2}{4\kappa[\kappa\cos^2(t\sqrt{\kappa})+\sin^2(t\sqrt{\kappa})]^2}.
\end{eqnarray}
It can be observed that the QFI indeed exhibits a growth scaling of $t^2$, which represents the Heisenberg scaling.

Using a similar scheme and experimental setup in Fig.\;\ref{circuit} and \ref{fig2}, we realize the non-unitary evolution operator $\hat{U}^\prime_\kappa=F\hat{U}_\kappa$, where $F$ is a real scalar.  It is important to note that the states $|\varphi^\prime_{H,V}\rangle$ realized in path $a$ and $b$ differ from those in the experiment of $\mathcal{PT}$-symmetry Hamiltonian, despite employing a similar experimental scheme and setup. In the previous experiment, the states in two paths are $|\varphi_H\rangle=|\psi_H\rangle/\sqrt{\langle\psi_H|\psi_H\rangle}$ and $|\varphi_V\rangle=|\psi_V\rangle/\sqrt{\langle\psi_V|\psi_V\rangle}$ respectively ($|\psi_H\rangle=\hat{U}_{PT}|H\rangle$, $|\psi_V\rangle=\hat{U}_{PT}|V\rangle$). However, in this experiment, we have  $|\varphi^\prime_{H}\rangle=|\psi_H^\prime\rangle/\sqrt{\langle\psi_H^\prime|\psi_H^\prime\rangle}$ and $|\varphi^\prime_{V}\rangle=|\psi_V^\prime\rangle/\sqrt{\langle\psi_V^\prime|\psi_V^\prime\rangle}$ ($|\psi^\prime_H\rangle=\hat{U}_\kappa|H\rangle$, $|\psi^\prime_V\rangle=\hat{U}_\kappa|V\rangle$). In this experiment, the total evolution operator is given by
\begin{eqnarray}
\hat{U}^\prime_{tot}&=&\gamma^\prime(|\psi^\prime_H\rangle\langle H|\otimes|a\rangle\langle a|+|\psi^\prime_V\rangle\langle V|\otimes|b\rangle\langle a|+|\psi_H^{\prime\perp}\rangle\langle V|\otimes|a\rangle\langle b|+|\psi_V^{\prime\perp}\rangle\langle H|\otimes|b\rangle\langle b|),\nonumber\\
&=&p^\prime|\varphi^\prime_H\rangle\langle H|\otimes|a\rangle\langle a|+q^\prime|\varphi^\prime_V\rangle\langle V|\otimes|b\rangle\langle a|+p^\prime|\varphi_H^{\prime\perp}\rangle\langle V|\otimes|a\rangle\langle b|+q^\prime|\varphi_V^{\prime\perp}\rangle\langle H|\otimes|b\rangle\langle b|,\nonumber\\
\end{eqnarray}
where $p^{\prime2}/q^{\prime2}=\langle\psi^\prime_H|\psi^\prime_H\rangle/\langle\psi^\prime_V|\psi^\prime_V\rangle$ and $\gamma^\prime=p^\prime/\sqrt{\langle\psi^\prime_H|\psi^\prime_H\rangle}=q^\prime/\sqrt{\langle\psi^\prime_V|\psi^\prime_V\rangle}$. By performing a projective operator $\hat{P}=I\otimes(|a\rangle+|b\rangle)(\langle a|+\langle b|)/2$, we obtain non-unitary evolution operator on probe qubit as follows
\begin{eqnarray}
\hat{U}^\prime_\kappa=(p^\prime|\varphi^\prime_H\rangle\langle H|+q^\prime|\varphi^\prime_V\rangle\langle V|)/\sqrt{2}=\gamma^\prime(|\psi^\prime_H\rangle\langle H|+|\psi^\prime_V\rangle\langle V|)=\gamma^\prime\hat{U}_\kappa,
\end{eqnarray}
where $|\varphi^\prime_H\rangle$ and $|\varphi^\prime_V\rangle$ are controlled by the angle of  H5, Q1, H6 and Q2. For each time point, we adjust these wave plates to ensure that the evolution operator is $\hat{U}^\prime_\kappa(t)$. To illustrate the distinctions between the two experiments, we present the angle configurations of H6 and Q2 at the same time points in both experiments (here, the probe state is $|0\rangle$, and it does not pass through H5 and Q1), as shown in Table.\;\ref{wphpt}.

\begin{figure}[h]
\centering
\includegraphics[ width=1\textwidth]{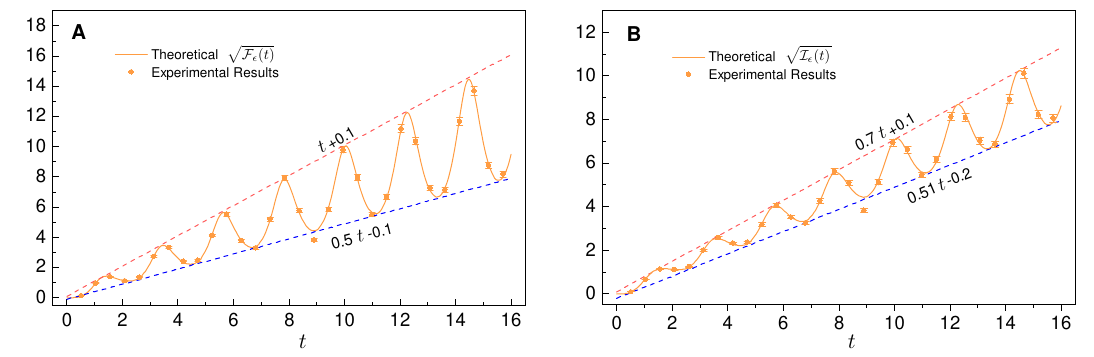}
\caption{QFI for varying time $t$. The probe state is set as $|0\rangle$, the measurement performed is $\hat{A}$, and the condition for optimal measurements is satisfied. The practical value of $\kappa$ we set is $2$. (A) The square root of QFI, the orange dots are the experimental data and the orange solid line is the theoretical value of $\sqrt{\mathcal{F}_\kappa(t)}$. (B) The QFI multiplied by normalized coefficient $K_\kappa$.}\label{NHL}
\end{figure}

We input the probe state $|\psi_0\rangle=|0\rangle$ and perform $n=1000\sim1200$ measurements $\hat{A}=|0\rangle\langle0|$ on output states for varying time $t$, where $\hat{A}$ is the corresponding optimal measurement for $|\psi_0\rangle$ according to the condition for optimal measurements Eq.\;(\ref{optmeas}),
\begin{eqnarray}
|f\rangle=\frac{-i[2t\sqrt{\kappa}-\sin(2t\sqrt{\kappa})]}{2\kappa\sin(2t\sqrt{\kappa})}|g\rangle.
\end{eqnarray}
The actual value of parameter is $\kappa=2$. For each time point, we repeat maximum likelihood estimation for 1000 times to obtain the distributions of the estimator $\hat{\kappa}$. As shown in Fig.\;\ref{NHL} , we present the estimation precision for varying time. The experimental results reveal that the precision follows Heisenberg scaling $t$ which aligns with the theoretical prediction. The exact experimental data is shown in Table.\;\ref{newtab}.

For this Hamiltonian without specific symmetries, the experimental results also match with the theory, which exhibits the generality of our theory.

\begin{table*}[h]
  \centering
  \normalsize
  \setlength{\tabcolsep}{2.5mm}
  \renewcommand\arraystretch{1.4}
  \begin{tabular}{c|c|cccccccc}
  \toprule[0.9pt]
  \multicolumn{2}{c}{${t}$} & {$\pi/2$} & {$\pi$} & {$3\pi/2$} & {$2\pi$} & {$5\pi/2$} & {$3\pi$} & {$7\pi/2$} & {$4\pi$}\\
  \hline
  \multirow{2}{1cm}{${\hat{H}_{PT}}$} & {H6} & {$-21.70^\circ$} & {$-40.21^\circ$} & {$83.56^\circ$} & {$66.00^\circ$} & {$42.61^\circ$} & {$-11.00^\circ$} & {$-26.33^\circ$} & ${-57.71^\circ}$\\
  \cline{2-10}
  & Q2 & $0^\circ$ & $0^\circ$ & $0^\circ$ & {$0^\circ$} & {$0^\circ$} & {$0^\circ$} & {$0^\circ$} & {$0^\circ$}  \\
  \hline
  \multirow{2}{1cm}{${\hat{H}_{\kappa}}$} & {H6} & {$111.45^\circ$} & {$55.67^\circ$} & {$-7.91^\circ$} & {$101.46^\circ$} & {$40.50^\circ$} & {$-17.01^\circ$} & {$93.21^\circ$} & {$26.39^\circ$}\\
  \cline{2-10}
  & {Q2} & {$0^\circ$} & {$0^\circ$} & {$0^\circ$} & {$0^\circ$} & {$0^\circ$} & {$0^\circ$} & {$0^\circ$} & {$0^\circ$}  \\
  \hline
  \bottomrule[0.9pt]
  \end{tabular}
  \caption{{The angle configurations of H6 and Q2 for varying times in two experiments.}}\label{wphpt}
\end{table*}

\begin{table*}[h]
  \centering
  \normalsize
  \setlength{\tabcolsep}{7pt}
  \renewcommand\arraystretch{1.5}
  \begin{tabular}{ccccc}\hline\toprule
  {$t$} & {Experiment $1/\sigma(\hat{\kappa})$} & {$\sqrt{\mathcal{F}_\kappa(t)}$} & {$E[\hat{\kappa}]$} & {error($\kappa$)} \\
  \hline
  {$\pi/6$}  & {$0.1139\pm0.0025$}  & {$0.1110$} & {$1.9841\pm0.2777$} & {-0.7938\%} \\

  {$2\pi/6$}  & {$0.9518\pm0.0213$}  & {$0.9762$} & {$1.9936\pm0.0332$} & {-0.3181\%} \\

  {$3\pi/6$}  & {$1.3815\pm0.0310$}  & {$1.3985$} & {$1.9955\pm0.0229$} & {-0.2260\%} \\

  {$4\pi/6$}  & {$1.1233\pm0.0251$}  &{$ 1.1274$} & {$1.9933\pm0.0282$} & {-0.3328\%} \\

  {$5\pi/6$}  & {$1.3494\pm0.0302$}  & {$1.3392$} & {$2.0172\pm0.0235$} & {0.8581\%} \\

  {$6\pi/6$}  & {$2.7372\pm0.0612$} & {$2.7642$} & {$2.0038\pm0.0116$} & {0.1882\%} \\

  {$7\pi/6$}  & {$3.3269\pm0.0744$}  & {$3.2765$} & {$1.9905\pm0.0095$} & {-0.4738\% }\\

  {$8\pi/6$}  & {$2.4127\pm0.0540$}  & {$2.3565$} & {$1.9863\pm0.0131$} & {-0.6873\%} \\

  {$9\pi/6$}  & {$2.4628\pm0.0551$}  & {$2.4002$} & {$1.9945\pm0.0128$} & {-0.2758\%} \\

  {$10\pi/6$}  & {$4.1355\pm0.0925$}  & {$4.1726$} & {$1.9983\pm0.0077$} & {-0.0841\%} \\
  \bottomrule
  \hline
  \end{tabular}
  \caption{{The standard deviations of $\hat{\kappa}$ for different times.}}\label{newtab}
\end{table*}

\end{document}